%% file: MSSM20.tex
\documentclass[12pt]{article}
\usepackage{graphicx}
\usepackage{amssymb}
\usepackage{amsmath}
\usepackage{longtable}
\usepackage{hyperref}

\input{distribution.inc}
\input{hiddenbranes.inc}

\input{famstatistics.inc}
\input{statistics.inc}

\def\beq{\begin{equation}}
\def\eeq{\end{equation}}
\def\bea{\begin{eqnarray}}
\def\eea{\end{eqnarray}}
\newcommand{\cA}{{\cal A}}

\newcommand{\cC}{{\cal C}}

\newcommand{\cG}{{\cal G}}
\newcommand{\cH}{{\cal H}}

\newcommand{\cK}{{\cal K}}

\newcommand{\cM}{{\cal M}}
\newcommand{\cN}{{\cal N}}

\newcommand{\cP}{{\cal P}}

\newcommand{\cR}{{\cal R}}
\newcommand{\cS}{{\cal S}}
\newcommand{\cT}{{\cal T}}

\newcommand{\cZ}{{\cal Z}}
\def\[{\left [}
\def\]{\right ]}
\def\({\left (}
\def\){\right )}
\newcommand{\R}[1]{(\ref{eq:#1})}
\newcommand{\ZZ}{\mathbb{Z}}
\newcommand{\ii}{{\rm i}}

\begin{document}

\vspace*{0.15in}
\hbox{\hskip 12cm NIKHEF/2004-015  \hfil}
\hbox{\hskip 12cm KUL-TF-04/35  \hfil}
\hbox{\hskip 12cm hep-th/0411129 \hfil}

\begin{center}
{\Large \bf Supersymmetric Standard Model Spectra from RCFT orientifolds\\}
\vspace*{0.5in}
{T.P.T. Dijkstra\footnote{email:tdykstra@nikhef.nl}${\,}$,
L. R. Huiszoon\footnote{email:lennaert\_janita@hotmail.com}${\,}$
and A.N. Schellekens\footnote{email:t58@nikhef.nl}${\,}$}\\
{\em
        NIKHEF Theory Group \\
	Kruislaan 409 \\
	1098 SJ Amsterdam \\
	The Netherlands \\
}

\end{center}

\begin{abstract}
We present supersymmetric, tadpole-free
$d=4,N=1$ orientifold vacua with a three family chiral fermion spectrum that is identical to that
of the Standard Model.
Starting with all simple current orientifolds of all Gepner models we
 perform a systematic search for such spectra. We consider several variations of
the standard four-stack intersecting brane realization of the standard model, with all quarks and leptons
realized as bifundamentals and perturbatively exact baryon and lepton number
symmetries, and with a $U(1)_Y$ vector boson that does not acquire a mass from Green-Schwarz terms.
 The number of supersymmetric Higgs pairs $H_1 + H_2$ is left free.
In order to cancel all tadpoles, we allow a ``hidden" gauge group, which must be
chirally decoupled from the standard model.
We also allow for non-chiral
  mirror-pairs of quarks and leptons, non-chiral exotics and 
  (possibly chiral) hidden, standard model singlet matter, as well
as a massless B-L vector boson. All of these less desirable features are absent in some cases, although not simultaneously.
In particular, we found cases with
massless Chan-Paton gauge bosons generating nothing more than $SU(3)\times SU(2) \times U(1)$.
We obtain almost 180000 rationally distinct solutions (not counting hidden sector
degrees of freedom), and present distributions of various quantities.
We analyse the tree level gauge couplings, and find a large range of values, remarkably centered around
the unification point.
\end{abstract}

\newpage

\section{Introduction}

String theory is hoped to be a consistent theory of quantum gravity, with the special feature that it strongly
constrains the matter it can couple to. Although direct experimental tests of new predictions seem out of reach
for the moment, it can at least be tested theoretically by verifying its internal consistency, in particular with
regard to gravity, and by checking that the limited set of matter it can couple to includes the standard model.
We may be lucky enough that the way the standard model is embedded in string theory implies predictions
for future experiments, but it may also happen that using all known experimental constraints we are still
left with more than one, or even a huge number of possibilities. But at present it is still a serious
challenge to find even one ``string vacuum" (which may actually be a metastable, approximate ground state)
that is a credible standard model candidate. To find such a vacuum requires an in-depth analysis of
known candidates based on robust criteria derived from experiment.  However, even
within the known classes of vacua, large areas have remained unexplored so far. As longs as that is the
case, nature would have to be very kind to us to allow us to find the ground state on which our universe is
based. In this paper we want to make a modest step towards broadening the set of accessible vacua, by means of
a systematic exploration of orientifolds of Gepner models.
It turns
out that this class is very rich, and includes an abundance of standard model-like spectra far
beyond anything that has come out of string theory so far. Some early, and partial
results were reported in \cite{Dijkstra:2004ym}.

Since 1984 there have been several attempts to search for standard model like string vacua.
Two general classes may be distinguished: heterotic string constructions
(either (2,2) or (0,2)), with the standard model particles realized as closed strings, and type-I (orientifold,
intersecting brane) constructions, characterized by open string realizations of the standard model.
Other possibilities certainly exist ({\it e.g.} M-theory compactifications), and since new ways of
obtaining the standard model are discovered about every five to ten years, it might be an illusion
to think that we are close to a complete picture.

To explore the two classes mentioned above, we have essentially three methods at our disposal:
free CFT constructions (free bosons and/or fermions and orbifolds thereof), geometric compactification
(in particular Calabi-Yau compactifications) and interacting CFT constructions. All three have
been applied successfully to heterotic string construction. The earliest example \cite{Greene:1986bm} of a three-family model
was based on the Tian-Yau Calabi-Yau manifold with Euler number 6 and a non-trivial homotopy group $\pi_1$ (as far as
we know still the only known manifold of this kind), used for Wilson line symmetry breaking. About a year later, the
first orbifold examples were found \cite{Ibanez:1987sn,Bailin:1987xm} (see also \cite{Font:1989aj} for later developments).
In \cite{Gepner:1987hi} a three-family model was constructed using tensor products of $N=2$ minimal models,
corresponding to a point in the moduli space of the Tian-Yau manifold. This model has an $E_6$ gauge group.
By reducing the $(2,2)$ world-sheet symmetry to $(0,2)$ a large number of related models was constructed in
\cite{Schellekens:1989wx} with $E_6$ or $SO(10)$ gauge symmetry. The symmetry in the bosonic sector can be reduced
further to obtain many models with a gauge group $SU(3)\times SU(2) \times U(1) \times [\hbox{Other factors}]$ and three families of
quarks and leptons \cite{Schellekens:1990unp}. Another class of three family models was obtained using free fermions
in \cite{Faraggi:1989ka,Faraggi:1991jr,Faraggi:1992fa}, and extended in \cite{Chaudhuri:1995ve}. By means of a
modified Calabi-Yau construction yielding $(0,2)$ spectra,
several more three-family models were obtained in\cite{Kachru:1995em}. All the foregoing models lead to level 1
realizations of the standard model gauge groups $SU(3)$ and $SU(2)$, and hence their spectra contain fractionally
charged states, or they have a non-standard normalization for the standard model $U(1)$ generator (or both)
\cite{Wen:1985qj,Athanasiu:1988uj,Schellekens:1989qb}. Unified models with standard model charge quantization
can be constructed by using higher level Kac-Moody algebras, and indeed three family models were obtained \cite{Kakushadze:1997mc}.
Yet another approach to getting the standard model, based on strongly coupled heterotic strings, was presented in \cite{Donagi:1999ez}.
All of these heterotic three family models are supersymmetric. Constructing non-supersymmetric heterotic strings is
much easier, but those theories are not stable, in general.

The open string road towards the standard model has been considerably more difficult. Until about 1995 this option
was rarely taken seriously, with a few exceptions \cite{Sagnotti:1987tw,Horava:1989vt}. This changed after the discovery of D-branes \cite{Po} and the observation in \cite{Witten:1996mz}
that the problematic relation between the unification scale and the Planck scale could be avoided in open string theories.
The construction of realistic theories is complicated by several new features, on top of those
of the closed type-II theory to which the orientifold procedure is applied: boundary and
crosscap states and the requirement of tadpole cancellation.
The first four-dimensional chiral, supersymmetric theory was constructed in
\cite{Angelantonj:1996uy}, but it was still far from realistic. In order to get standard model-like
spectra,
in several later papers ({\it e.g.}
\cite{BlGKL0007,AlFIRU0011,IbMR0105,BlKLO0107,BaKL0108,BlBKL0206,Kokorelis:2002sm,Kokorelis:2003jr,CaU0303})
the requirement of supersymmetry was relaxed, and only part of the the tadpoles
were cancelled, namely the RR tadpoles needed for consistency. The resulting theories are not stable, but
otherwise consistent. The first semi-realistic supersymmetric theory that satisfies all tadpole conditions
 was presented in \cite{CvSU0107} and \cite{Cvetic:2001tj,CvP0303}.
However, these spectra contains chiral exotics, in addition to the standard model representations.
In \cite{BlGO0211,H0303} a three family supersymmetric
Pati-Salam ($SU(4)\times SU(2) \times SU(2)$) model was constructed, but the two $SU(2)$s do not emerge
directly from string theory but from a diagonal subgroup of Chan-Paton groups using a ``brane recombination"
mechanism. In \cite{CvLL0403} supersymmetric Pati-Salam models were found that emerge directly from the
CP-groups, but with additional chiral exotic matter, and in \cite{CvPS0212} supersymmetric $SU(5)$ GUT
models with chiral exotics were presented. Most of the foregoing constructions are based on
orientifolds of  of toroidal orbifolds \cite{BiSa,GiP}, except for \cite{BlBKL0206} which uses the quintic Calabi-Yau manifold.
In \cite{Cascales:2003wn} standard model like spectra where obtained using branes at singularities.
The first investigation of string spectra from open strings of orientifolds of Gepner models was done
in \cite{comments}, in 6 dimensions. The first analysis in four dimensions was done by \cite{BlW9806}. These authors did not find chiral spectra, but got a first glimpse of the vast
landscape of solutions. Further work on this kind of construction, including some
chiral spectra, was presented in
\cite{BlW9806,GoM0306,AlALN0307,Bl0310,BrHHW0401,BlW0401,Aldazabal:2004by}. In \cite{Dijkstra:2004ym},
the precursor of the present paper,
the first examples were found of supersymmetric standard model spectra, with the standard model group
appearing directly as a Chan-Paton group and without chiral exotics. Shortly thereafter the first such
example was found in  the context of orientifolds of toroidal orbifolds \cite{Honecker:2004kb}. In 
\cite{Marchesano:2004xz,Marchesano:2004yq} the first semi-realistic examples were found with flux comapctifications and
partially stabilized moduli, although with chiral exotics.
For recent results on chiral fermions from Gepner orientifolds see \cite{Aldazabal:2004by}. For recent reviews
with more references consult for example \cite{Ur0301,Ot0309,K0310,Lu0401}.

Recently there has been a lot of work on (non)-supersymmetric vacua with moduli stabilized
by three-form fluxes on a Calabi-Yau manifold. Although this is an extremely interesting development,
we will not consider it here, simply because the methods we use do not allow this at present. 

For a variety of reasons we will require the string spectrum to be supersymmetric. The first reason
is phenomenological.
Although we do not commit ourselves
to a supersymmetry breaking mechanism or scale,
the most obvious scenario is the standard one: supersymmetry
breaking at a few TeV, induced by gaugino condensation in a hidden sector (which exists in most of our models), with
supersymmetry playing the role of protecting the gauge hierarchy. Indeed, such a hierarchy inevitably exist in these
models, since six dimensions are compactified on a Calabi-Yau manifold at a rational point in moduli space, and
hence there is no reason to expect some of the compactified dimensions to be extremely large.
There are two other reasons why we want to keep supersymmetry unbroken. First of all,
we can then be certain that the four-dimensional
strings we construct are stable and consistent. But the
most important reason is a practical one. Space-time supersymmetry has the effect of extending
the world-sheet chiral algebra, thereby organizing the fields into a smaller number of primaries.
This is what makes
our computations manageable in practice.

The use of rational conformal field theory (RCFT) in these constructions has well-known advantages and
disadvantages.
The advantage of the algebraic approach is that we can explore a large class of models with uniform methods.
But clearly the disadvantage is that one ends up in a special point in moduli space, both with
regard to the Calabi-Yau manifold, as well as the choice of branes wrapping it. It is not reasonable to
expect such a ground state to be exactly the standard model, because many observable quantities, such as
quark and lepton masses and gauge couplings will depend on the moduli, which
have been fixed at a specific value. Clearly
we should focus on those features that do not depend on the moduli. The primary feature to consider is then of course
the chiral spectrum.

Apart from  supersymmetry breaking there are several other important issues that we do not consider here, such as
standard model symmetry breaking, Yukawa couplings, Majorana masses for the neutrinos, etc.
We see our results therefore mainly as a first exploration of some interesting regions in the huge landscape
of possible models. Once a promising type of model has been identified, one may try to explore it in more detail,
either by CFT perturbations in the neighborhood of the special point, or by constructing the corresponding
Calabi-Yau and the set of branes on it, using the RCFT results as a guideline. Even within the context of RCFT
one could
push these models further and compute certain couplings, but unfortunately
the required CFT techniques are not yet available for all couplings. For example,
three-point couplings between open strings are in principle computable in RCFT, but to develop this formalism to
include non-trivial modular invariant partition functions and simple current fixed points would still require
a substantial amount of work. Gauge couplings, on the other hand, are easily computable, and we do so in all cases.

Since we end up with a very large set of solutions, our results should give a reasonable idea of what
kind of spectra one may expect, and one can perform some statistical analysis on this set, somewhat
similar in spirit (though with a quite different philosophy) to the approach presented in \cite{Douglas:2003um}.
In addition, despite the inherent limitations of the algebraic approach,
one may explore the effect of brane moduli as well as some Calabi-Yau moduli. The former, since for a
given CY-manifold we typically find a large number of spectra, which can be interpreted in terms of
branes in different discrete positions; the latter, because some distinct Gepner points may lie
on the same Calabi-Yau moduli space. Especially the brane position moduli seem to be probed rather
effectively by rational points, in certain cases.

\subsection{Brane configurations considered}

In this paper we consider boundary states in a rational type IIB CFT. The relevant physical open string
quantities are annulus and Moebius coefficients. For the sake of clarity, it is convenient to
present the models in terms of an intersecting brane picture, although such a picture is not
really used in our construction. This picture would be appropriate in a large volume limit and
for type IIA string, the mirror of what we consider here. The geometric interpretation of the construction
considered here is presumably in terms of magnetized D3 and D7 branes \cite{BlGKL0007}. In the following,
by the ``intersection" of two branes a and b we mean $\sum_i A^i_{ab} \chi^i(\tau/2)|_0$, where
$A^i_{~ab}$ is an annulus coefficient and $\chi$ 
is the character restricted to massless states. By the ``chiral intersection" we mean
the same quantity restricted to chiral states.

We consider here a specific type of intersecting brane models, based on a four-stack configuration with
a baryon brane, a weak brane (or left brane), a right brane and a lepton brane, labeled a,b,c,d
respectively \cite{IbMR0105}. These are the minimal brane configurations with baryon and lepton number conservation
and all quarks and leptons realized as bifundamentals. The Chan-Paton gauge groups associated with
these branes contain the standard model gauge group.
In addition we allow ``hidden branes", with gauge groups with respect to which all
standard model particles are neutral. These branes were introduced to cancel massless tadpoles, but their
gauge groups may play a useful phenomenological r\^ole, in particular for gaugino condensation.

The a and d branes are required to be complex, and carry
Chan-Paton group $U(3)_{\rm a}$ and $U(1)_{\rm d}$ respectively. The former contains the standard model
gauge group $SU(3)$. The overall phase factor of $U(3)_{\rm a}$ corresponds to baryon number, and the
$U(1)_{\rm d}$ to lepton number. In the standard model these symmetries are not gauged, and anomalous.
In string theory these anomalies are canceled by Green-Schwarz terms, involving a bilinear
coupling of the ``bary-photon" and ``lepto-photon" to massless two-form fields. These couplings
produce a mass for the linear combination $B+L$ of these $U(1)$ bosons, breaking the corresponding
combination of baryon and lepton number. Nevertheless the broken symmetry still prevents the
appearance of dangerous baryon and lepton number violating couplings at least perturbatively \cite{IbMR0105}.

The b and c branes may be real or complex. In the standard four-stack realization of the standard model
the first family emerges as follows if they are both complex, with CP-groups $U(2)_{\rm b}$ and $U(1)_{\rm c}$ respectively
\bea
(u,d): & [{\rm a},{\rm b}] \hbox{ or } [{\rm a},{\rm b}*] \cr
u^c: & [{\rm a}^*,{\rm c}] \cr
d^c: & [{\rm a}^*,{\rm c}^*] \cr
(e^-,\nu): & [{\rm b},{\rm d}] \hbox{ or } [{\rm b}^*,{\rm d}] \cr
e^+ & [{\rm c},{\rm d}^*] \cr
\nu^c: & [{\rm c}^*,{\rm d}^*] \cr
\eea
Here $[x,y]$ denote strings beginning on brane $x$ and ending on brane $y$, and $x^*$ is the brane
conjugate to $x$.
The $Y$-charge of the standard model is given by the linear combination
$$ Y = \frac16 Q_{\rm a} - \frac12 Q_{\rm c} -\frac12 Q_{\rm d} \ .$$ The overall phase symmetry in $U(2)_{\rm b}$ is always anomalous
with respect to the $a$ and the $d$ branes and the corresponding gauge boson acquires a mass;
the surviving gauge group is $SU(2)_W$.  Note that $B+L$ and $U(2)_{\rm b}$ have independent anomalies with respect
to the standard model, so that there is no linear combination of these phases symmetries that remains unbroken.

The standard model weak gauge group can also be constructed out
of real branes on top of an orientifold plane, yielding $Sp(2)$. Since the spectrum is real with respect to
the c-brane, we may allow the group $O(2)_{\rm c}$ here instead of $U(1)_{\rm c}$, with $Q_{\rm c}$ replaced by the
(properly normalized) $O(2)$
generator in the definition of the charge. Since $O(2)$ branes differ from
$Sp(2)$ branes only by a Moebius sign, we decided to allow the latter as well.
Strictly speaking this is a departure from our philosophy of looking only for the
simplest standard model realizations, but these models are as easy to look for as
$O(2)$ models, and have the interesting feature of yielding
``left-right symmetric" models with an $SU(2)_L \times SU(2)_R$ gauge group.
Such
gauge groups appear as part of the symmetry-breaking chain of the Pati-Salam model, and indeed in some examples there
are related spectra with the $d$ brane on top of the $a$ brane, yielding precisely the Pati-Salam model.
Then we end up with following six types of models:
\begin{eqnarray*}
\hbox{Type 0} & \; U(3) \times Sp(2) \times U(1) \times U(1) \\
\hbox{Type 1} & \; U(3) \times U(2) \times U(1) \times U(1)   \\
\hbox{Type 2} & \; U(3) \times Sp(2) \times O(2) \times U(1)  \\
\hbox{Type 3} & \; U(3) \times U(2) \times O(2) \times U(1)  \\
\hbox{Type 4} & \; U(3) \times Sp(2) \times Sp(2) \times U(1)  \\
\hbox{Type 5} & \; U(3) \times U(2) \times Sp(2) \times U(1)   \\
\end{eqnarray*}

The complete spectrum of these theories can contain massless vector bosons from three sources: the standard model
part of the open sector, as listed above, additional ``hidden" branes, and the closed sector. The
latter gauge bosons are nearly inevitable, but do not have minimal couplings
to the quarks and leptons. The gauge bosons from the hidden sector may be absent altogether, but in
any case do not couple to the standard model particles. If we ignore these two kinds of vector bosons, the gauge group is
quite close to that of the standard model. As a Lie-algebra, it is a non-abelian extension of the standard model only
for types 4 and 5; in all other cases we get the standard model with at most one additional non-anomalous $U(1)$ factor, $B-L$.

The gauge bosons coupling to the non-anomalous symmetries $Y$ and $B-L$ can acquire a mass from Green-Schwarz type
two-point couplings to two-form fields, provided that these couplings do not generate a contribution to the anomaly.
We find that in most cases such a mass contribution is absent, and that $Y$ is more likely to acquire a mass than $B-L$.
The latter statement is based only on the models presented in \cite{Dijkstra:2004ym}, where
masslessness of $Y$ was only used as an {\it a posteriori} check. In the present paper brane stacks yielding non-zero
$Y$-mass were eliminated before attempting to solve the tadpole equations, but no condition was imposed on the $B-L$ mass.
We found a massive $B-L$ photon in about $\percMasslessBminL\%$ of the type-0 models, and for none of the type-1 models (of which we found only very few). There are even models with a massive $B-L$ and
no extra branes at all. These models (\masslessBminLandNoBranes in total) have precisely the standard model gauge group from the open sector,
but there are still 18 additional Ramond-Ramond vector bosons from the closed sector.

We expect masslessness of the Y-boson, in addition to lepton and quark chirality and supersymmetry, to be
among the features of these models that are unaffected by generic perturbations around the rational point.
The potential origin of a Y-boson mass would be the generation of a two-point coupling to one of the RR two-forms away
from the rational point.
However, such a two-point coupling
would very likely generate an anomalous contribution in combination with a
three-point coupling of the same RR-two form
to two gauge bosons.

The same logic applies to the $B-L$ gauge boson. If it is massless at the string level,
it should acquire a mass trough a fundamental or dynamical Higgs mechanism,
just as the $Z$ and $W$ bosons. Candidates for the required Higgses are the sneutrinos or two standard model/
hidden sector bifundamentals (one ending on the c-brane and one on the d-brane), bound by a hidden sector gauge group.

Most other features cannot be expected to survive generic perturbations. In particular this concerns
the massless particles that are non-chiral
with respect to $G_{\rm S}$, namely the mirrors, rank-2 tensors, leptoquarks and standard model/hidden sector bifundamentals.
It seems plausible that for many of them masslessness is an artifact of being in a rational
point in moduli space. They will get a mass when the moduli are changed, and one can investigate this by computing
the coupling of the closed string moduli to the massless fermions.
However, some of the massless particles correspond to brane position moduli, and hence to
flat directions in the superpotential. One of the problems we have in investigating this and other more detailed phenomenological
issues is that we have to find a way to do a meaningful analysis for a huge number of solutions.

There are many other brane configurations that would yield standard-model-like spectra. For example
another attractive possibility might be to start with a $U(5)$ stack (realizing an $SU(5)$ GUT)
plus other branes or the reduction
of such a stack to $U(3) \times U(2)$. In such models some quarks and leptons emerge as anti-symmetric tensors
and baryon and lepton number are not symmetries. Models of this type were studied in \cite{CvPS0212}, but so far
with the disappointing result that there are additional chiral symmetric tensors of $SU(5)$. In principle one could search
for models of this kind in exactly the same way, just as one could search for Pati-Salam type models.
It goes without saying that if we were to relax some constraints and allow for example chiral exotics or diagonal
embeddings of standard model factors in the CP group, then the number of solution would almost certainly explode.
Nevertheless such options, although unattractive, are not necessarily ruled out experimentally.

\subsection{Chirality}

Let us explain more precisely what we mean by ``getting the standard model from string theory", since
this issue tends to cause confusion. Quite generally,
one might accept any string spectrum if the gauge group $G_{\rm S}$
emerging directly from open or closed strings contains
$G_{\rm SM}=SU(3)\times SU(2) \times U(1)$, and if the fermion representation of $G_{\rm S}$ reduces to
three times $(3,2,\frac16)+(3^*,1,\frac13)+(3^*,1,-\frac23)+(1,2,-\frac12)+(1,1,1)$, written in
terms of left-handed Weyl fermions. These should be the only fermions that are chiral with
respect $SU(3)\times SU(2) \times U(1)$. If at this stage there were
additional chiral particles one could still imagine mechanisms that give them a sufficiently large
mass after standard model symmetry breaking and $SU(3)$ confinement, but
if that were true we simply need more experimental input to go ahead.
There may be additional
massless fermions that are non-chiral with respect to $SU(3)\times SU(2) \times U(1)$
and their may be additional fermions, chiral with respect to $G_{\rm S}$, that become non-chiral after the
reduction from $G_{\rm S}$ to $SU(3)\times SU(2) \times U(1)$.
    
The group $G_{\rm S}$ is the complete CP-group from the open sector times any gauge group
generated by closed sector vector bosons. The physical realization of the group theoretical 
``reduction" mentioned above can take many forms, such as
mass generation for $U(1)$'s by Green-Schwarz anomaly cancelling terms, confinement or breaking by
a fundamental or dynamical Higgs effect. Furthermore part of $G_{\rm S}$ may remain unbroken,
if the corresponding gauge bosons do not couple to quarks and leptons. We did not commit ourselves
here to a particular reduction mechanism. In the majority of the spectra we consider, $G_{\rm S}$
is embedded in $G_{\rm SM}$ as $G_{\rm S}=G_{\rm SM} \oplus G_{\rm Hidden}$ (as a Lie-algebra),
except in types 4 and 5, where the $U(1)_{\rm c}$ group is non-trivially extended to $Sp(2)$.
In the latter cases an additional Higgs-like mechanism would be required to arrive at the
standard model. Just as is the case with the supersymmetry breaking and the standard model
Higgs mechanism, one could impose additional constraints on the results in order for a
a particular mechanism to be realized, but such constraints are less robust and more model-dependent
than the requirement of chirality, our main guiding principle.

The possibilities for
$G_{\rm S}$-chiral particles that are $G_{\rm SM}$-non-chiral are the following
\begin{enumerate}
\item Right-handed neutrinos. These particles are singlets (and hence not chiral)
w.r.t. $G_{SM}$ but are chiral with
respect to lepton number, which is broken. In our case, there are always three of them. This is
an inevitable consequence of tadpole cancellation, which cancels the cubic anomalies in $U(1)_{\rm d}$, plus
the fact that we only allow bifundamentals of the (a,b,c,d) branes to be chiral.
\item Higgsinos. In the MSSM the fermionic partners of the Higgs are non-chiral with respect to $G_{SM}$, but
in models of types 1, 3 and 5 there is a possibility for the Higgses to be chiral with respect to $U(2)_{\rm b}$.
This gauge symmetry breaks to $SU(2)$ in the first step, but its initial presence can still forbid
the generation of large masses for the Higgs.
This is a desirable feature, as it may give a mechanism for getting light Higgs bosons.
\item Mirror quarks and leptons, which are chiral with
respect to $U(2)_{\rm b}$. These particles can appear for the same reason as the Higgsinos, but are less desirable. For
example the $U(3) \oplus U(2)$ combinations $(3,2)+(3^*,2)$ is chiral, but becomes non-chiral when $U(2)_{\rm b}$
is reduced to $SU(2)$. We have allowed such particles in principle, but (just as the chiral Higgsinos), they
occur only rarely.
\item $G_{SM}$ singlets, which are chiral with respect to the hidden gauge group. Such particles couple only
the the SM-matter gravitationally, and hence are acceptable as ``dark matter", if not too abundant. Furthermore they may
acquire a mass and/or be confined by $G_{\rm Hidden}$ dynamics.
\end{enumerate}

Unwanted chiral matter within the standard model sector can be avoided by the selection of  a,b,c  and  d
branes, and chiral matter from open strings stretching between the SM branes and the hidden branes can
be avoided by appropriate selection of the latter. One could in principle also forbid chiral rank-2
tensors within $G_{\rm Hidden}$ by an {\it a priori} constraint,
but it is very hard to forbid chiral bifundamentals
within $G_{\rm Hidden}$, except by constructing all solutions and eliminating them {\it a posteriori}.
Since constructing
all solutions is nearly impossible in most cases, we decided to allow $G_{\rm SM}$-singlets
that are chiral with respect to $G_{\rm Hidden}$.
Nevertheless, they occur in only a small fraction (about 12.5\%) of our solutions.
This is largely due to the fact that
our search is biased in favor of few additional branes, and it is harder to make chiral spectra with
fewer branes, and impossible with a single brane.

\subsection{Scope of the search}\label{sec:models_tp_alg}

In this paper we consider all modular invariant partition functions (MIPFs) that are
symmetric simple current modifications of the charge conjugation
invariant of all 168 minimal N=2 tensor products.
The precise number of such MIPFs is determined as follows.
Generically, it is just a matter of applying the procedure of \cite{GaS92,KrS} and restricting
to symmetric bi-homorphisms $X$ (defined more precisely in \cite{KrS} and in the next chapter). However,
if there
are identical N=2 factors in the tensor product, there will be equivalences among these MIPFs, and
we remove equivalent ones. Furthermore, for small values of $k$
 (the ``level" of the minimal
model), especially $k=2$, generically distinct simple current invariants are in fact identical, and these are
also removed from the set. We then end up with 5403 distinct MIPFs. They can be characterized
in part by the resulting Hodge numbers $h_{11}$ and $h_{21}$, and by the number of
gauge singlets in the heterotic string spectrum. These numbers can be compared to
tables of such spectra produced about fifteen years ago \cite{Schellekens:1989wx,Fuchs:1990yv}.
Unfortunately a complete comparison
is difficult, because the old results are either no longer available,
and certainly not in electronic form, or the search 
was not fully exhaustive, or the symmetry of the MIPFs was not specified. But to the extent
that a comparison is possible the results seem to agree.

In a few cases there appear to be further equivalences,
or at least some MIPFs may
correspond to distinct rational points on the same Calabi-Yau space. In total we found
\totBettis\ different  combinations of Hodge numbers, and \totHeterotic\ Hodge numbers plus singlets
({\it i.e} gauge singlets in the Heterotic spectrum.)

For each of these MIPFs we consider all orientifolds allowed by the general formula of \cite{FHSSW}.
These orientifolds are subject to three equivalence relations, one originating from
permutations of identical factors, and two as part of the general formalism. These
equivalences are removed, and we then end up with a total of 49322 {\it a priori} distinct
orientifolds. Our results indicate that indeed they are generically distinct.

For each MIPF and orientifold we consider the complete set of boundaries. This means
that the number of boundaries is equal to the number of Ishibashi states of the MIPF.
The MIPF may be of any type: pure fusion rule automorphisms, extensions of the chiral
algebra, or combinations thereof. The beauty of the
formalism of \cite{FHSSW} is that it works independently of such details.
It is well-known that in the case of extensions one
can distinguish boundaries that respect the extended symmetry and boundaries that do not.
A complete set of boundaries contains both kinds. The CFT we start with is itself an
extension of the minimal model tensor product CFT (by alignment currents and the
space-time supersymmetry current). Those extensions are respected by all our solutions, by construction.
If the MIPF extends the chiral algebra further, one could work directly in the extended
CFT and only consider boundaries that respect the extension. One would then find a subset
of our solutions. Alternatively, one could also start with less symmetry, and treat
for example space-time supersymmetry as a bulk extension. This would allow, in principle,
supersymmetry breaking boundaries. In practice this is quite
hard, because the number of primary fields increases dramatically. Undoubtedly, so will
the number solutions.

Our goal was to complete this analysis for all MIPFs, in order to arrive
at a picture that is as complete as possible.
Unfortunately, the analysis could not be completed in all cases. Two tensor combinations had
too many primaries to finish the computation of chiral intersections in a reasonable amount of time.
In five others the number of candidate (a,b,c,d) branes was so large that we decided to
restrict ourselves to types 0 and 1, of which there are fewer. For a given MIPF, orientifold
and type, the number of four-stack candidates was more than a million in some cases.

The main computational stumbling block are the tadpole equations. For every valid set of (a,b,c,d)
branes, the number of variables is equal to the number of boundaries that do not have a chiral
intersection with a,b,c and d. This number can become as large as several hundreds, for a few tens
of tadpole equations.
Obviously, the time
needed to evaluate this completely grows exponentially with the surplus of variables over
equations.
This means that beyond a certain number of variables it is impossible to decide conclusively
that there are no solutions. In those cases we did perform a systematic search for
solutions with 0,1 and 2 hidden branes, and 3 if the number of variables was less than 100,
4 if the number was less than 400. Furthermore, in the simpler cases we attempted solving the
equations in general. Of the \totMipfs~MIPFs, \absentMipfs~were not analysed at all, and \mipfsWmissTypes~only for types 0 and 1,
and in \mipfsWaborted~cases the tadpole equations were not fully analysed
(\mipfsWsolsWaborted~of these did have solutions, however).

There are several possible ways to count solutions as distinct. On the one hand, one could
regard solutions as identical if they are connected to each other by continuous, non-singular
variations of open or closed string moduli. In an RCFT approach this is hard to do, since we
cannot vary the moduli in a continuous way. The other extreme would be to count all distinct
massless spectra, including the hidden sector gauge groups and representations. This is also
not possible in our case, since we did not do a systematic search through all hidden sector gauge
groups. We have chosen an intermediate criterium: solutions are regarded as distinct if they are of
different type or have a different massless (chiral and non-chiral) standard model spectrum.
Furthermore we treat different dilaton couplings for the a,b,c,d branes and the O-plane as a
distinction, and the absence or presence of a hidden sector. By contrast, in \cite{Dijkstra:2004ym}
hidden sector distinctions were also counted. The number of solutions for one of the MIPFs of the
tensor product $(6,6,6,6)$ quoted in that paper (``more than 6000") reduces to 820 with our
present way of counting.

\subsection{Contents}

This paper is organized as follows. In the next chapter, we review the 
ingredients of the algebraic orientifold construction. 
Section \ref{sec:spectrum} contains a general discussion of the massless 
spectrum. In section \ref{sec:tad_anom} we discuss tadpole and anomaly 
cancellation. Chapter \ref{sec:results} contains our results. Due to the huge 
number of solutions it is impossible and pointless to present
detailed spectra. Therefore we only give distributions of several quantities of interest, such as the number of Higgs scalars. We also analyse the values of 
gauge coupling ratios at the string scale. In section \ref{sec:simplest} we 
present one example in more detail, a model without any additional branes that 
in several respects is the simplest we encountered. In 
section \ref{sec:conclusions} we formulate some conclusions.   
An essential computational technique
is to organize
the Ishibashi and boundary labels into simple current orbits, which leads to a dramatic 
speed-up of the calculations. This is explained in the appendix.

\section{Algebraic Model Building}\label{sec:modelbuilding}

Our starting point is four-dimensional type-II string obtained by tensoring
NSR fermions with a combination of N=2 minimal models with total central charge 9.
In a covariant description, the NSR part of the theory is built out of four fermions $\psi^{\mu}$
with Minkowski metric, and a set of superconformal ghosts. The type-II theory is modular invariant
and has two world-sheet supersymmetries needed for consistency. We assume it to be symmetric in left-
and right-moving modes and have an extended chiral algebra leading to two space-time supersymmetries.
To this theory we apply the orientifold procedure. Since our approach is based on unitary rational
CFT, it is convenient to use this description not only for the minimal N=2 factors, but also for
the NSR part of the theory. To do so we use a bosonic description of the latter (see \cite{LeSW} and
references therein). This is convenient because model-independent
complications due to the GSO projection, spin-statistics and superconformal ghosts are automatically taken
care of. This implies that we are formally constructing bosonic open strings. To obtain the spectrum we
mimic the procedure used in the covariant approach, namely fix a ghost charge to select the physical states. This
translates into a consistent truncation of the bosonic string characters. In the case of closed strings, this
procedure can be shown to map modular invariant bosonic strings to modular invariant fermionic strings. In the
case of open strings, it leads to fermionic open strings satisfying all the integrality conditions on torus
and Klein bottle as well as Annulus and M\"obius strip amplitudes, and that have the correct spin-statistics for
all physical states, and the proper symmetrization for Ramond-Ramond states.
We emphasize that this is only used here
as a bookkeeping device, and that we are not trying to conjecture a relation between fermionic and bosonic
strings.

Our starting point is a class of bosonic string theories with chiral algebra
\beq \label{eq:alg1}
E_{8} \otimes D_{5} \otimes {\cal A}_{int} \;\;\; ,
\eeq
where $E_8$ and $D_5$ are level 1 affine Lie algebras models.
In this paper we can take the
 model-dependent factor $\cA_{int}$ to be of $4d$ Gepner type
\beq
\cA_{int} = \otimes_{k=1}^r {\cal A}_k \;\;\; , \;\;\; c_{int} = 9
\eeq
and $\cA_k$ is the $N=2$ minimal model at level $k$. The $E_8$ factor
has no influence on the massless spectrum.
The only role of this factor is to cancel the conformal anomaly
 $c_8 + c_5 + c_{int} + c_{bos} + c_{ghosts} = 8 + 5 + 9 + 4 - 26 =0$ where $c_{bos}$ is the
contribution of the uncompactified bosons. The $D_5$ factor describes the lightcone NSR fermions plus the
longitudinal NSR fermions and superconformal ghosts. The truncation that we implement at the end of
the day basically amounts to removing the contribution of the latter.  For the construction of the
type-II string we follow the procedure explained in \cite{Schellekens:1989wx} (see also \cite{Fuchs:2000gv})
for heterotic strings, the
only difference being that the fermionic truncation is applied to both the left- and the rightmoving sector. 
The tensor product (\ref{eq:alg1}) is first extended by means of {\it alignment currents} (even
combinations of the world-sheet supercurrents of the factors), needed to maintain $N=1$ world-sheet
supersymmetry. To get theories with $N=2$ space-time supersymmetry we must extend the algebra~\R{alg1} further
by the simple current group generated by the
currents $(0,S,S,S,S,...)$, where $S$ is a spinor representation of $D_5$ or a Ramond ground state of
each minimal model that is a simple current (in $D_5$ and each of the minimal models there are two choices,
but which one we take is irrelevant as long as the same choice is made in both chiral sectors. For the
minimal models we choose $(l,q,s)=(0,1,1)$, in the usual notation).

Because all aforementioned chiral algebra extensions
are of simple current type, all chiral data like the spectrum of primaries
$\{i\}$, the conformal weights $h_i$, the modular matrix $S_{ij}$ and the fixed point resolution matrices
$S^J$ can be expressed in terms of the chiral data of the original unextended tensor product.

In this paper use a left-right symmetric extension in order to be able to apply
the boundary/crosscap state formalism of \cite{FHSSW}.  There is a second, asymmetric choice obtained by using
instead of $(0,S,S,S,S,...)$ the simple current $(0,C,S,S,S,...)$ in of the chiral sector.
These are called type IIB (symmetric) and type
IIA (asymmetric) extensions respectively.
                                             
After applying these extension one obtains
168 \cite{Lutken:1988zj} four-dimensional type-IIB theories. Most of these have
$N=2$ space-time supersymmetry, which will be broken to $N=1$ by the orientifold procedure. Five of the 
168 theories have $N=4$ supersymmetry, and can be ignored for further purposes, as they will never yield
chiral $N=1$ open strings. We treat all these theories as non-supersymmetric CFTs. From the world-sheet
point of view, the world-sheet supercurrents are fields with conformal weight $\frac{3}{2}$ which are
not in the chiral algebra (although their even powers are), and the space-time
supercurrents are extended chiral algebra currents with conformal weight 1, which are treated just as any
other extension. There is one reason why conformal weight 1 currents are special, and that is that the
sub-algebra they generate is an affine Lie-algebra or $U(1)$ factor. In this case they extend $D_5$ to $E_6$
(or $E_7$ in the $N=4$ theories). In this way we obtain and RCFT with $N_{\rm prim}$ primaries, whose
ground states are in some representation of $E_6$, of the general form
$m_0 ({\bf 1}) \oplus m^+ ({\bf 27}) \oplus m^- ({\bf 27}^*)$. For the $N=2$ theories, $N_{prim}$ varies between 260 and 108612.

These CFTs are our starting point and their chiral algebras are left unbroken in the rest of the procedure.  
We will refer to it as the {\it susy chiral algebra}.
As was discussed in \cite{Fuchs:2000gv}, one could consider the possibility of starting with a smaller chiral
algebra, and allow for the possibility that -- for example -- space-time supersymmetry is present in the bulk, but
broken on some of the branes. While it is possible in principle to investigate this in our formalism, the practical
problem is that the chiral algebra becomes smaller, and hence the number of primaries much larger.

Among the $N_{prim}$ primaries there is almost always a subset $N_{sim}$ that are simple currents. This number
ranges from 2187 (equal to $N_{prim}$) for the tensor product $(1)^9$ to just 1.
Under fusion, they form a discrete group $\cG$.
These simple currents are
used to build symmetric modular invariant partition functions.

For a simple current MIPF one has to specify the following data
\begin{itemize}
	\item A group $\cH$ that consists of simple currents of $\cA$. All currents $J$ in $\cH$ must 
 satisfy the condition that the product of their conformal weight $h_J$ and order $N_J$ is integer. In
	general $\cH$ is a product of cyclic factors $\cH = \prod_{\alpha} \ZZ_{N_\alpha}$. The generator of the
	$\ZZ_{N_\alpha}$ will be denoted as $J_\alpha$. 
	\item A symmetric matrix $X_{\alpha\beta}$ that obeys
\bea
2 X_{\alpha\beta} & = & Q_{J_\alpha}(J_\beta) \mod 1 , \alpha\neq \beta \\
 X_{\alpha\alpha} & =&  -h_{J_\alpha}
\eea
	plus a further constraint $N_\alpha X_{\alpha\beta} \in \ZZ$ for all $\alpha,\beta$.
\end{itemize}       
Here $Q$ is the simple current monodromy charge, $Q_J(a)=h(a)+h(J)-h(Ja)$, where $h$ is
the conformal weight.
When in the following we write $X(J,J')$ for
	arbitrary simple currents in $\cH$ we mean
\beq
X(J,J') = \prod_{\alpha,\beta} n_\alpha m_\beta X_{\alpha\beta}
\eeq
	for $J=\prod_\alpha J_\alpha^{n_\alpha}$ and $J'=\prod_\alpha J_\alpha^{m_\alpha}$.

The resulting value of $Z_{ij}$ is the number of currents $L\in \cH$ such that
\bea
j & = & Li \\
 Q_M(i) + X(M,L) & = & 0 \mod 1 \;\;\;
\eea
	for all $M \in \cH$.
These MIPFs can be further chiral algebra extensions of the susy chiral algebra, 
fusion rule automorphisms or combinations thereof.
The formalism of \cite{FHSSW} is insensitive to the distinction among these various types.
With the
exception of certain pathological cases, this set of MIPFs is the most general one where the 
combinations $(i,j)$ of left and right representations that occur are linked by simple currents,
i.e. $i=Jj^c$ for some $J\in cH$.  As the ``$c$" indicates, we build simple currents starting from
the charge conjugation invariant. One could also start from the diagonal invariant, but there is no
general formula available for the boundary and crosscap coefficients in that case. In many (though not
all) cases the diagonal invariant is itself a simple current automorphism of the charge conjugation invariant,
and hence is already included. It is well-known that additional, ``exceptional" MIPFs exist for the 
Gepner models (see \cite{Schellekens:1990jh}, \cite{Gannon:1994sp}), 
including the famous ``three-generation" one for the tensor product $(1,16,16,16)$ \cite{Gepner:1987hi}, but
again there is no boundary/crosscap formalism available for these cases (although the boundary coefficient are known
for the $SU(2)$ exceptional invariants \cite{Behrend:1999bn}).

The next step is to specify the orientifold data, which consist of
\begin{itemize}
	\item A {\em Klein bottle current} $K$. This can be any simple current of $\cA$ that is
	 local with all order two currents in $\cH$, i.e., obeys 
\begin{equation} \label{KBCcond}
Q_I(K)=0\mod1 \;\forall I\in \cH, I^2=0.
\end{equation}
\item A set of phases $\beta_K(J)$ for all $J\in \cH$ that satisfy
\beq 
\beta_K(J) \beta_K(J') = \beta_K(JJ') e^{2\pi\ii X(J,J')}  \;\;\; , J,J' \in \cH
\eeq
	with $\beta_K(J) = e^{i \pi ( h_{KL}-h_K)} \eta(K,L)$, with $\eta(K,L)=\pm 1$.
\end{itemize}  
Note that the phases $\beta_K(J)$ satisfy the same multiplication rule independent of the Klein bottle current $K$,
and that the solutions depend on $K$ because of the second requirement. The signs $\eta$ can be chosen freely provided
the multiplication rule for $\beta$ holds. It is easy to see that this implies that there is a freedom of choosing one
sign for each independent even factor in the simple current subgroup $cH$. All these choices yield valid orientifolds,
but they are some equivalences, which will be discussed below.

This data defines a (bosonic) orientifold with spectrum encoded in the total one-loop partition function
\beq \label{eq:Part}
\frac12\(\cT + \cK + \cA + \cM\)
\eeq
where we distinguish between the four topologically distinct surfaces with vanishing
Euler number. These contributions can be expanded in (bi)linears of (hatted) characters in the usual way \cite{AnS}:
\bea \label{eq:torus}
\cT = \sum_{ij} Z_{ij} \chi_i \chi_{j}^* & & \;\;\; , \;\;\; \cK = \sum_i K^i \chi_i \\
\cA = \sum_{ab} \cN_{a} \cN_{b} A^{i}_{~ab} \chi_i 
& & \;\;\; , \;\;\; \cM = \sum_{a} \cN_{a} M^i_{~a} \hat{\chi}_i \;\;\; .
\eea
The following objects are introduced:
\begin{itemize}
	\item The labels $a,b$ that appear in the open string sector of the partition function are a
short-hand notation for the {\em boundary labels}. In full glory these labels are $\cH$-orbits $[a]$ of a chiral sector $a$ with a possible
 degeneracy label $\psi_a$ which is a (discrete group) character of the a certain subgroup of the
stabilizer, called the
{\it central stabilizer} $\cC_a$ (see \cite{FHSSW} and below). We write this as $[a,\psi_a]$.
	\item The nonnegative integers $\cN_a := \cN_{[a,\psi_a]}$ are the CP-factors. These numbers must be
	such that the total partition function is free of divergences. This will be reviewed in 
	section~\ref{sec-tadpole}. 
\end{itemize}

The {\em Klein bottle}, {\em annulus} and {\em M\"obius coefficients} factorize as
\bea 
K^i & = & \sum_{m,J,J'} \frac{ S^i_{~m}U_{(m,J)} g^{\Omega,m}_{J,J'} U_{(m,J')} }{S_{0m}} \label{Kcoef} \\
A^{i}_{[a,\psi_a][b,\psi_b]} & = & \sum_{m,J,J'}
	\frac{ S^i_{~m}R_{[a,\psi_a](m,J)} g^{\Omega,m}_{J,J'} R_{[b,\psi_b](m,J')} }{S_{0m}} \label{Acoef} \\
M^i_{[a,\psi_a]} &  = & \sum_{m,J,J'} \frac{ P^i_{~m}R_{[a,\psi_a](m,J)} g^{\Omega,m}_{J,J'} U_{(m,J')} }{S_{0m}}
\label{Mcoef}
\eea
where $P=\sqrt{T}ST^2S\sqrt{T}$~\cite{P} and $S,T$ are the usual modular matrices. In these expressions 
the sums run over all {\em Ishibashi labels}. These labels are pairs $(m,J)$ that obey
\begin{eqnarray} \label{eq:Ishi}
m & = & Jm \;\;\;,\\
Q_K(m) + X(M,J) &= & 0 \mod 1
\end{eqnarray}
for all $M \in \cH$. We note that we consider boundaries and crosscaps of ``trivial automorphism type"  \cite{ReS}, which
means that we require that the susy chiral algebra is preserved (and not just preserved up to automorphism) in
closed string scattering from one of these defects. This implies that
the closed strings that can couple to these defects must be C-diagonal.
The Ishibashi labels~\R{Ishi} are one-to-one to such
closed string sectors. We have also introduced
\begin{itemize}
	\item The {\em Ishibashi metric} $g^{\Omega,m}$
\beq
g^{\Omega,m}_{J,J'} =  \frac{S_{m0}}{S_{mK}} \beta_K(J) \delta_{J',J^c} \;\;\; .
\eeq
for all $J,J'\in\cH$. Here $\Omega$ indicates the choice of Klein bottle current as well as the
phases $\beta_K(J)$ that define an orientifold.
	\item The {\em boundary reflection coefficients} 
\beq \label{eq:BRC}
R_{[a,\psi_a](m,J)}  =  \sqrt{\frac{|\cH|}{|\cC_a||\cS_a|}}
 \psi_a^*(J) S^J_{am}
\eeq 
	\item The {\em crosscap reflection coefficients}
\beq  \label{eq:URC}
 U_{(m,J)}  =  \frac{1}{\sqrt{|\cH |}} \sum_{L\in \cH}
\eta(K,L) P_{LK,m}\delta_{J,0}  \;\;\;  \;\;\; .
\eeq
\end{itemize}
where $S^J$ is the
{\em fixed point resolution matrix} $S^J$, whose rows and columns are labelled by fixed points $a,m$ o
	f $J$,
	implements a modular $S$-transformation
	 on the  torus with $J$ inserted.
	 It is unitary and obeys \cite{FSS}
\beq \label{eq:FPact}
S^J_{Ki,j} = F_i(K,J) e^{2\pi\ii Q_K(j)} S^J_{ij} \;\;\; .
\eeq
The aforementioned {\em central stabilizer} is defined in terms of this quantity as
\beq
\cC_a = \{J \in \cS_a | F_a(K,J)e^{2\pi\ii X(K,J)} = 1 \;\; \hbox{for all} \;\; K \in \cS_a\} .
\eeq

One can check that $R$ is unitary.
The reflection coefficients have an important
physical meaning, because they are (proportional to) the coupling of closed strings from Ishibashi sector $(m,J)$
to D-brane $[a,\psi_a]$. The {\em oriented} annulus amplitude therefore reads
\beq
[A^{\rm or}]^{i\;\;\; [b,\psi_b]}_{[a,\psi_a]} := \sum_{m,J,J'}
	\frac{ S^i_{~m}R_{[a,\psi_a](m,J)} R_{[b,\psi_b](m,J')}^* }{S_{0m}}
\eeq
In the unoriented string, specified by the Klein bottle current $K$ and the phases $\beta(J)$, the annulus
amplitude is
\beq
A^{i}_{[a,\psi_a][b,\psi_b]} = [A^{\rm or}]^{i\;\;\; [b,\psi_b]^c}_{[a,\psi_a]}
\eeq
where $[b,\psi_b]^c$ is the conjugate boundary label. Geometrically the pair of branes $[b,\psi_b]$ and
$[b,\psi_b]^c$ are mapped to each other by the orientifold action $\Omega \cR = \Omega \cR(K,\beta)$.
In CFT this image is encoded by the boundary conjugation matrix
\beq
A^{0}_{[a,\psi_a][b,\psi_b]} = \left\{ \begin{array}{ll}
					1 &, [b,\psi_b] = [a,\psi_a]^c \\
					0 &, \hbox{otherwise}
					\end{array}
\right.
\eeq
Note that the various unoriented annuli can all be obtained from the unique oriented annulus by
multiplication with the boundary conjugation matrix. 

The physical meaning of $U$ is as coupling constants between the
Ishibashi sectors $(m,J)$ and the O-plane. 

This formalism has been shown to lead to integer values for all open and
closed string particle multiplicities \cite{thesis_lennaert}. This results holds
for all RCFT's, not just the minimal N=2 models considered here. This universal validity
gives additional confidence in its correctness, but the ultimate consistency check would be to
demonstrate that all sewing constraints are satisfied on all Riemann surfaces. This has
been done for orientable surfaces \cite{Fuchs:2004dz}, and is underway for the non-orientable case.

\subsection{Orientifold equivalences}

For a general simple current MIPF the set of known orientifolds is parametrized
by a Klein bottle current $K$ and a number of signs $\epsilon$, one for each independent even
factor in the discrete group that defines the MIPF. The Klein bottle current can
be any simple current subject to the constraint (\ref{KBCcond}), and there is no restriction
on the signs. However, not all these choices are inequivalent. The following
equivalences exist between the choices $\{K,\epsilon\}$
(here ${\cal G}$ is the full group of simple currents and ${\cal H}$
the subgroup used in the construction of the MIPF)
$$\{K,\epsilon\} \sim \{KJ^2,\epsilon'\},\ \  J \in {\cal G}$$
$$\{K,\epsilon\} \sim \{KL,\epsilon''\},\ \  L \in {\cal H}$$
$$\{K,\epsilon\} \sim \{\pi(K),\ \ \hat\pi(\epsilon)\}$$
Here $\pi$ is the action induced by the permutation of identical minimal
models (if any) on the primary fields of the tensor product, and $\hat\pi$ is
the action induced on the signs $\epsilon$. The modified signs $\epsilon'$ and
$\epsilon''$ can be worked out from the formula for the crosscap coefficients
(\ref{eq:URC}) and the relation
$$ P_{J^2a,b} = \epsilon_{J^2}(a)e^{2\pi i [Q_J(b) - Q_J(Ja)]} P_{ab}\ ; \ \ \ \epsilon_J(a)=e^{\pi i[h_a-h_{Ja}]}\ , $$
but we will not present them explicitly here. The combined action of the three
equivalences organizes the various crosscap choices into equivalence classes, and
we have taken into account one representative from each class. The results seem to
indicate that there are no further equivalences: in general the number of (a,b,c,d) stacks
of various types, as well as the number of tadpole solutions is distinct.

Note that in each case the equivalence between orientifolds holds up to a certain
permutation of the boundary labels. A subgroup of these transformation may fix the
orientifold, but lead to a residual equivalence of boundary choices for a given orientifold.
 We did not attempt
to remove this (and other) equivalence among boundaries, because it was much easier to
compare the resulting spectra and remove identical ones {\it a posteriori}.

\section{Massless Spectrum}\label{sec:spectrum}

Our prime interest will be the identification of the massless states in the partition function~\R{Part}. Before
doing this in the correct way, we have to perform a truncation to obtain the spectrum of the
superstring. This truncation is most easily described as follows. Note
 that the current $(0_8,S_5,\vec{S})$ has spin $1$ and that therefore the bosonic 
 algebra $\cA$ must contain a level one WZW factor that is larger than $D_{5}$.
 The only possibility is $E_{6}$. All primaries $m$ of $\cA$
 therefore decompose into primaries of $E_{6}$. Tachyonic states are always singlets of $E_6$.
 Massless states are singlets, fundamentals ${\bf 27}$, anti-fundamentals ${\bf 27}^*$ or adjoints
 ${\bf 78}$. The truncation from the bosonic spectrum to a superstring spectrum is
\bea 
\hbox{left-movers} 	& & {\bf 1} \rightarrow - \;\;\;  \nonumber \\
			& & {\bf 27} \rightarrow \frac12 \Psi \;\;\; , \;\;\; {\bf 27}^* \rightarrow \frac12 \Psi^* \nonumber\\
			& & {\bf 78} \rightarrow V \;\;\;  \nonumber\\
\hbox{right-movers} 	& & {\bf 1} \rightarrow - \;\;\;  \nonumber\\
			& & {\bf 27} \rightarrow \frac12 \Psi \;\;\; , \;\;\; {\bf 27}^* \rightarrow \frac12 \Psi^* \label{eq:PSSR1}\\
			& & {\bf 78} \rightarrow V \;\;\;  \nonumber
\eea
where $\Psi$ is a (complex) $N=1$ chiral multiplet and $V$ a $N=1$ vector multiplet. Note that $E_6$ singlets are
 projected out. The $\bf 27$ representation thus yields one real bosonic degree of freedom, and one
fermionic one. In heterotic strings a representation $(\bf 27, R)$ (where $R$ is
some gauge representation) is always
accompanied by a $(\bf 27, R^*)$, and together they form one $N=1$ chiral multiplet, containing
a complex boson and a Weyl fermion in the representation $R$.
In type-II closed strings, the combinations $(\bf 27,\bf 27)+(\bf 27^*,\bf 27^*)$ yields one
N=2 vector multiplet, with four real bosonic and four real fermionic degrees of freedom. The
combination $(\bf 27,\bf 27^*)+(\bf 27^*,\bf 27)$ yields one $N=2$ hyper multiplet, with
the same number of degrees of freedom. Note that in principle one could switch the r\^ole of
the $\bf 27$ and the $\bf 27^*$ in the truncations for the right-movers with respect to the left-movers.
In covariant
language, this corresponds to switching the ghost-charge assignment for the fermions. This would
map a IIA spectrum to a IIB spectrum and vice-versa. However, the same interchange can
also be achieved by going from a IIB to a IIA extension. In applications to orientifolds,
it is clearly preferable to adopt a universal truncation rule (ghost charge assignment) for
left and right-moving, as well as open string characters.

  To every $h_i=1$ primary we can associate a Witten index
 \beq
 w_i = m^+_i - m^-_i 
 \eeq
where $m^+_i$ ($m^-_i$) counts the number of ${\bf 27}$ (${\bf 27}^*$) in $i$.  Note that
\beq
m^\pm_{i^c} = m^\mp_{i} \rightarrow w_{i^c} = - w_i
\eeq

\subsection{The Oriented Closed String Spectrum}

The torus contribution $\cZ$ in~\R{Part} is the partition function of the parent theory of the orientifold. 
After undoing the bosonic string map it describes type II string theory on
some Calabi-Yau $3$-fold at the Gepner point. 
The ground state of the vacuum sector $(00)$ is projected out. At the first excited
level it yields 
\beq
V * V = G + H \;\;\; ,
\eeq
the $N=2$ gravity and universal hyper multiplet. The other massless sectors
 yield the model-dependent Abelian vector and  
hyper multiplets. First note that modular invariance implies 
\beq
Z_{ij} = Z_{i^cj^c}
\eeq
Complex sectors $(ij) + (i^cj^c)$ contribute the following $N=2$ multiplets
$$ Z_{ij} \(m^+_im^+_{j} + m^-_im^-_{j}\) \;\;\; \hbox{vector multiplets} $$
$$ Z_{ij} \(m^+_im^-_{j} + m^-_im^+_{j}\) \;\;\; \hbox{hyper multiplets} $$
Real sectors $(ij)$, $i=i^c,j=j^c$ contribute
$$ Z_{ij} m^+_im^+_{j} \;\;\; \hbox{vector multiplets} $$
$$ Z_{ij} m^+_im^+_{j} \;\;\; \hbox{hyper multiplets} $$
The total numbers $h_{21}$ of vector multiplets and $h_{11}$ of
hyper multiplets are
\bea
h_{21} & = & \frac12 \sum_{ij} Z_{ij} \(m^+_im^+_{j} + m^-_im^-_{j} \) \\
h_{11} & = & \frac12 \sum_{ij} Z_{ij} \(m^+_im^-_{j} + m^-_im^+_{j} \)
\;\;\; .
\eea
The sum is over all fields, including conjugates,
and the factor $\frac12$ corrects for double-counting; for real
fields $m^+_i=m^-_i$. Note that
\beq
\chi := 2[h_{21} - h_{11}] = \sum_{ij} w_i w_{j^c} Z_{ij}
\eeq
is a topological invariant. The string theory we have constructed therefore has the same massless spectrum
as type IIB string theory on a Calabi-Yau manifold $X_3$ with Hodge numbers $h_{11}$ and
$h_{21}$\rlap.\footnote{Note that in \cite{Dijkstra:2004ym} the Hodge numbers of the type IIA compactification were
listed. In this paper we list them for the type IIB compactification,  the
closed string theory to which we apply the orientifold procedure.}

The spectrum of the other dual pair, $\(\text{IIA}/{X}_3,\text{IIB}/\tilde{X}_3\)$, can be obtained by conjugating the
right-moving space-time supercurrent, which results in the IIA extension.
It is easy to see that there are $h_{21}$ hyper multiplets and $h_{11}$ vector multiplets in this case.

\subsection{The Unoriented Closed String Spectrum}

The first step in the orientifold procedure is the truncation of the type II
spectrum to states that are invariant under the involution $\Omega=\Omega(K,\beta)$.
The resulting massless $d=4,N=1$ spectrum can be obtained from $\frac12 (\cZ +
\cK)$ by simple counting arguments. The vacuum sector gives the universal
gravity multiplet and a chiral multiplet that contains the dilaton.
Off-diagonal sectors do not flow in the direct Klein bottle. Their contribution is halved by the
orientifold projection. Complex off-diagonal sectors $(ij) + (i^cj^c) + (ji) + (j^ci^c)$ contribute
the following $N=1$ multiplets
$$ Z_{ij} \(m^+_im^+_{j} + m^-_im^-_{j}\) \;\;\; \hbox{vector multiplets} $$
$$ Z_{ij} \(m^+_im^+_{j} + m^-_im^-_{j}\)
+ 2Z_{ij} \(m^+_im^-_{j} + m^-_im^+_{j}\)\;\;\; \hbox{chiral} \;{\rm  multiplets} $$
Real off-diagonal sectors $(ij)+(ji)$ contribute
$$ Z_{ij} m^+_im^+_{j}  \;\;\; \hbox{vector multiplets} $$
$$ 3Z_{ij} m^+_im^+_{j} \;\;\; \hbox{chiral multiplets} $$
Diagonal sectors are symmetrized or anti-symmetrized according to the Klein bottle coefficient.
Complex diagonal sectors $(ii) + (i^ci^c)$ contribute
$$ \frac12( Z_{ii} - K_i) \(m^+_im^+_{i} + m^-_im^-_{i}\) \;\;\; \hbox{vector multiplets} $$
$$ \frac12( Z_{ii} + K_i) \(m^+_im^+_{i} + m^-_im^-_{i}\) + Z_{ii} \(m^+_im^-_{i}
+ m^-_im^+_{i}\)\;\;\; \hbox{chiral multiplets} $$
Real diagonal sectors contribute
$$ \frac12( Z_{ii} - K_i) m^+_im^+_{i} \;\;\; \hbox{vector multiplets} $$
$$ \frac12( Z_{ii} + K_i) m^+_im^+_{i}  + Z_{ii} m^+_im^-_{i}
\;\;\; \hbox{chiral multiplets} $$
Define
\beq
h_{11}^\pm  :=  \frac14 \[ \sum_{ij} \( Z_{ij} \pm  \delta_{ij} K_i \)
\(m^+_im^+_{j} + m^-_im^-_{j}\) \] ,
\eeq
where we sum over all primaries. Then the total number of closed string
Abelian
vector multiplets is $h_{11}^-$ and the total number of model-dependent
closed string chiral multiplets is $h_{21} + h_{11}^+$. In a geometrical
setting the numbers $h_{11}^+$ and $h_{11}^-$ denote the number of harmonic
$(1,1)$-forms that are anti-invariant or invariant under the orientifold
action.

\subsection{The Oriented Open String Spectrum}

The massless gauge bosons of the $\Pi_a U(N_a)$ gauge group
come from the first excited level of the vacuum sector. 
The annulus coefficient $A^{i~b}_{~a}$ counts states in the
bifundamental $(V_a, V^*_b)$ representations of the
space-time gauge group. Note that
\beq ~\label{eq:CPTan}
A^{i^c~b}_{~a} = A^{i~a}_{~b} \;\;\; .
\eeq
Let $M_{a}^{+~b}$ ($M_{a}^{-~b}$) denote the number of  chiral (anti-chiral) 
multiplets that transform according to $(V_a, V^*_b)$. These numbers are given by
\beq
M_{a}^{\pm~b} = \sum_i m_i^\pm A^{i~a}_{~b} \;\;\; 
\eeq
where we sum over all primaries. Due to~\R{CPTan} the spectrum obeys the $d=4$ CPT relation $M_{b}^{\pm~a}
= M_{a}^{\mp~b}$. The net chirality is measured by the anti-symmetric chiral intersection matrix
\beq
I_a^{~b} = M_{a}^{+~b} - M_{a}^{-~b} = \sum_i w_i A^{i~b}_{~a} \;\; .
\eeq

\subsection{The Unoriented Open String Spectrum}

The open string spectrum of the orientifold is encoded in $\frac12 (\cA + \cM)$. 
The boundary conjugation matrix $A_{~ab}^0$ defines the orientifold image or 
{\em conjugate} $a^c$ of brane $a$. 
Complex pairs of
branes $a \neq a^c$ give rise to unitary gauge groups. For real branes $a=a^c$ the gauge group 
depends on the M\"obius coefficient $M_a^0$. When $M_a^0=-1/+1$, the first excited level of the vacuum
is symmetrized/anti-symmetrized, signalling a $Sp(N_a)/SO(N_a)$ gauge group. 
We can summarize this as
\beq
G = \otimes_{a, {\rm complex}} U(N_a) \otimes_{a,{\rm real}} SO(N_a) \otimes_{a,{\rm pseudo-real}} Sp(N_a)
\eeq
The CP-factors $N_a$ are determined by tadpole cancellation (see subsection~\ref{sec-tadpole}).
In our conventions $A^i_{ab}$ counts states in $(V_a, V_b^*)$. When we conjugate a brane label, the
corresponding vector must be conjugated. So $A^i_{a^cb}$ counts states in $(V_a^*,V_b^*)$ etcetera.
Note that\footnote{This follows from
\beq
R^*_{a(m,J)} = g^{\Omega,m}_{JJ'} R_{a^c(m,J')} \;\;\;    
\eeq
which can be derived from $A^i_{~ab} = A^{i~b^c}_{~a}$, unitarity of $S$ and completeness.
} 
\beq \label{eq:CPTunor}
A^{i^c}_{ab} = A^{i}_{a^cb^c} \;\;\; .
\eeq
For off-diagonal sectors $a \neq b$, the sectors $(a,b)$ and $(b,a)$ must be identified. The total number
$M^+_{ab}$ ($M^-_{ab}$) of chiral (anti-chiral) 
multiplets that transform according to $(V_a, V_b^*)$ is
\beq
M^\pm_{ab} = \sum_i m_i^\pm A^i_{ab} \;\;\; 
\eeq
where we sum over all primaries. Due to~\R{CPTunor} the spectrum obeys the $d=4$ CPT relation
$M^\pm_{a^cb^c} = M^\mp_{ab}$. The net chirality of fermions transforming as $(V_a,V_b^*)$ is measured by
\beq
\Delta_{ab} = M^+_{ab} - M^-_{ab} = \sum_i w_i A^i_{~ab} \;\; .
\eeq
Note that the ordering of the indices is irrelevant. From~\R{CPTunor} we easily derive
\beq \label{eq:symDelta}
\Delta_{ab} = -\Delta_{a^cb^c} \;\;\; , \;\;\;  \Delta_{ab^c} = -\Delta_{a^cb}
\eeq
When $b=a^c$ the representations are adjoints (Adj) of $U(N_a)$.
 Of course, adjoint matter cannot give rise to
chiral matter in $d=4$, as can easily be seen from~\R{symDelta}. In order to make contact with geometry, 
we note that $\Delta_{ab}$ is the upper-half part of the geometric 
intersection matrix and that the lower-half is given by $\Delta_{a^cb^c}$. 
Diagonal sectors $a=b$ are projected by the M\"obius strip to symmetric (S) or anti-symmetric (A)
representations of the gauge group $G(N_a)$. Note that~\footnote{We now also need
\beq
U^*_{(m,J)} = g^{\Omega,m}_{JJ'} U_{(m,J')} \;\;\; \;\;\; .   
\eeq
}
\beq \label{eq:CPTmo}
M^{i^c}_{~a} = M^i_{~a^c} 
\eeq
In a self-explanatory notation, the number of chiral (anti-chiral)
multiplets in these representations are
\bea
M^\pm_{a,{\rm S}} & = & \frac12 \sum_i m^\pm_i (A^i_{aa} + M^i_a)\\
M^\pm_{a,{\rm A}} & = & \frac12 \sum_i m^\pm_i (A^i_{aa} - M^i_a)\;\;\; .
\eea 
From~\R{CPTunor} and~\R{CPTmo} these multiplicities obey the CPT relations $M^\pm_{a,{\rm S}} 
= M^\mp_{a^c,{\rm S}}$ and $M^\pm_{a,{\rm A}} = M^\mp_{a^c,{\rm A}}$. The net
chirality is 
\bea
\Delta_{a,{\rm S}} & = & \frac12 \sum_i w_i (A^i_{aa} + M^i_a) \\
\Delta_{a,{\rm A}} & = & \frac12 \sum_i w_i (A^i_{aa} - M^i_a) \;\;\; .
\eea 
For real branes $a$, this index vanishes, as befits the symmetric and anti-symmetric
representations of symplectic and orthogonal groups in $d=4$. When we compare with a
 geometric description, $\sum_i w_i A^i_{aa}$
is the intersection between a brane and its image, whereas $\sum_i w_i M^i_{a}$ is the intersection between a
brane and the orientifold plane(s). 

\section{Tadpoles \& Anomalies}\label{sec:tad_anom}

\subsection{Tadpole cancellation} \label{sec-tadpole}

Non-vanishing one-point functions of massless scalars on the disk or crosscap
may cause several problems. If the scalar is physical particle in the spectrum, surviving
the orientifold projection, a tadpole indicates an instability in the vacuum. This
manifests itself as an infinity in the Euler number 0 diagrams. In this case
the theory would be unstable, but one might still hope that a stable vacuum exists.
 If the scalar is not a physical particle, the presence of a tadpole renders the theory
inconsistent, and this may manifest itself through an uncanceled anomaly. If all Klein
bottle coefficients are positive, all scalars from NS-NS sectors are physical, but
since the R-R-sector always has a projection with opposite sign, the R-R-scalars are
unphysical.

In a supersymmetric theory the NS-NS and R-R sectors are linked, and canceling
unphysical tadpoles is equivalent to cancelling all tadpoles.  The condition
for the cancellation of all tadpoles is
\beq \label{eq:tadpole}
\sum_{b} N_b R_{b(m,J)} = 4 \eta_m U_{(m,J)}
\eeq
for all Ishibashi labels $(m,J)$ for which the sector $(mm^c) + (m^cm)$ in the torus \eqref{eq:torus} yields massless
space-time scalars. Here $\eta_0=1$ and $-1$ otherwise. Tadpole cancellation is a condition on the CP-factors 
$N_a$ of the gauge groups.

There are two further constraints on the CP multiplicities. If two boundaries $a$ and $b$ are conjugate,
one must require that $N_a=N_b$, and if the CP-group associated with label $a$ is symplectic
$N_a$ must be even.

The dilaton couplings $R_{0b}$ are always positive (the Ishibashi label $m=0$ is non-degenerate, so
there is no need for the degeneracy label $J$). Hence one can only satisfy the dilaton tadpole
condition if $U_0 < 0$. The overall sign of the crosscap coefficients is a free parameter,
which must be fixed so that $U_0 < 0$. Changing this sign changes the sign of all M\"obius
coefficients.

\subsection{Anomaly cancellation}

The chiral gauge anomalies can be obtained from a formal polynomial that
is proportional to
\beq \label{eq:POL}
\cP(F) =  \sum_{i} w_i \[\sum_{ab} A^i_{ab} {\rm ch}_a (F) {\rm ch}_b (F) + \sum_a M^i_a {\rm ch}_a (2F)\]
\eeq
where
\beq
{\rm ch}_a (F) := \sum_n \frac{1}{n!} {\rm Tr}_{a} F^n \;\;\; , \;\;\;
\eeq
where the trace in
${\rm Tr}_{a} F^n $ is taken over the fundamental
representation of $U(N_a)$, and over the anti-fundamental in ${\rm Tr}_{a^c} F^n$; $F$ is the
field strength two-form.
To obtain the cubic anomalies in four dimensions one restricts to polynomial to six-forms and applies
the descent method. The argument $F$ can be expanded in a Lie-algebra basis with generators $T^k$ as $\sum_k F^k T^k$. If
$F$ lies entirely within a real subalgebra all odd terms (in $F$) in the polynomial vanish, and in particular there are
no four-dimensional anomalies.
 
Following ~\cite{BiM}, we can
we can easily show that tadpole cancellation implies
cancellation of the purely cubic terms in the polynomial, but not the others (the
``purely cubic" terms are obtained by keeping only terms proportional to ${\rm Tr} F^3$, without using
group-dependent factorizations of such terms (such factorizations of cubic traces exist only for $U(1)$ and $U(2)$).
To do so, we use that fact that the chiral characters or Witten indices
transform to themselves under modular transformations, as well as under transformations involving the $P$-matrix, because
they are independent of the modulus $\tau$
\def\Tr{{\rm Tr~}}
\beq \label{eq:SPWitten}
w_i = i \sum_j S^i_j w_j \;\;\; , \;\;\; w_i = i \sum_j P^i_j w_j
\eeq 
substituting this into the cubic part of the polynomial, and using the expression for the annulus and M\"obius strip
amplitudes (\ref{Acoef}) and  (\ref{Mcoef})  we get
\bea
 \cP(F)_{\rm cubic} =  &-i \sum_{m,J,J'}  (S_{0m})^{-1} w_m \sum_{a} R_{(m,J)a}g^{\Omega,m}_{J,J'} \\
  &\times \left\{\sum_b R_{(m,J')b}
           \left [N_a \Tr_b F^3 + N_b \Tr_a F^3 \right]
   +U_{(m,J')}
           \left[8 \Tr_a F^3 \right]\right\}
\eea    
Now we interchange the summed indices $a$ and $b$ in the second term, use the fact that $g^{\Omega,m}_{J,J'}= g^{\Omega,m}_{J',J}$,
and substitute for $U_{(m,J)}$ the righthand side of (\ref{eq:tadpole}) (note that only terms with $m\not=0$ contribute, because
the vacuum sector is non-chiral in four dimensions), and we see immediately that $\cP(F)_{\rm cubic}=0$. 

For non-abelian factors in the gauge group this implies simply the usual cancellation of cubic anomalies. For $U(1)$-factors that
get chiral contributions only from vectors, it also has the expected consequence: the total number of vectors and
conjugate vectors must be the same in each such factor.
The situation is a bit more interesting as soon as chiral tensors
contribute.  For example, if we assign charge $\pm 1$ to the (anti)-fundamental representations of a $U(1)$ factor, then
symmetric and anti-symmetric tensors have charge $\pm2$. The anomaly cancellation implied by tadpole
cancellation has nothing to do with the third power of these charges, but is an extrapolation of $U(N)$ anomaly cancellation to 
$N=1$. Hence we get a contribution
proportional to $\pm 1$ for vectors,
$\pm(N-4)=\mp 3$ for anti-symmetric tensors and $\pm(N+4)=\pm 5$ for symmetric ones
(note furthermore that anti-symmetric $U(1)$ tensors do not even exist; in a ``massless" anti-symmetric sector the first state in the
spectrum is a massive symmetric tensor).
This is not a problem, because
precisely for $U(1)$ factors the cubic traces factorizes into lower order traces, which are not cancelled by
tadpole cancellation anyway, but are removed by the Green-Schwarz mechanism.

In this paper we only encounter vector/anti-vector anomaly cancellation within the standard model gauge groups because
we have required that all chiral particles attached to the (a,b,c,d) branes should be bifundamentals. 
This implies
in particular the presence of three left-handed anti-neutrinos to cancel the c and d-brane anomalies. It might be worth
considering to drop the restriction to bifundamentals for the c and d branes and achieve anomaly cancellation by
means of tensors, but we will not pursue that possibility here.

In those cases where the b-brane group is $U(2)$, tadpole cancellation imposes the constraint that the numbers
of vectors and anti-vectors of $U(2)$ should be equal. This constraint is discussed in \cite{IbMR0105}, and
lead in that context to a quantization of the number of families in multiples of three.  This is not the case
here, because the Higgs also makes contributions to the $U(2)$ anomaly.
In the hidden sector more interesting examples of anomaly cancellation are possible, because chiral tensors
may (and indeed do) contribute. These anomaly cancellations are of course a useful check on our results, 
but they are still far less restrictive than
in six dimensions, where we have performed such checks on the complete solution to the tadpole solutions a few years ago.

The surviving part of the anomaly polynomial,
\beq
\cP (F)_{\rm rest} =  \sum_{i} w_i \sum_{ab} A^i_{ab} \left[{\rm Tr}_{a} F {\rm Tr}_{b} F^2  +  {\rm Tr}_{a} F^2 {\rm Tr}_{b} F\right]
\eeq 
must be cancelled by a generalized Green-Schwarz mechanism. 
This mechanism involves coupling between RR$p$-forms
$C_{(p)}$ and the $U(N_a)$ field strength $F_a$. For the oriented open string these couplings are
\beq~\label{eq:CS}
\sum_a S_a = \sum_a \int_{M_4} C^a_{(2)} N_a F_a + C^a_{(0)} {\rm Tr}_a F^2
\eeq
where we decided to expand the RR fields in the highly degenerate basis spanned by the complete set of boundary
labels $a$. Here $F_a$ is the $U(1)_a$ field strength and ${\rm Tr}_a F^2$ is a trace over the vector
representation of $U(N_a)$. In a geometric language the branes wrap homology classes $\pi_a$ and we can choose
a basis $\omega_a$ of
$3$-forms such that $\int_{\pi_a} \omega_b = \delta_{ab}$. The RR forms are KK
reductions along this basis. We have $\int_{CY} \omega_a \wedge \omega_b = I_{ab}$, the
chiral intersection matrix of branes $a$ and $b$. Poincare duality in this basis then 
reads 
$$\star {\bf d}C^a_{(2)} = \sum_b I_{ab} {\bf d}C^b_{(0)}\;\;\; .$$
From the couplings~\R{CS} we can easily read off 
the contribution to the $U(1)_aSU(N_b)^2$ mixed anomaly,
$$ N_a I_{ab} \;\;\; .$$
For unoriented strings the sectors $a$ and $a^c$ are identified and the couplings are encoded in
\bea
\sum_a \(S_a + S_{a^c}\) & = & \sum_a \int_{M_4} N_a \[ C^a_{(2)} -  C^{a^c}_{(2)}\] F_a + \\
	& &  + \sum_a \int_{M_4} \[ C^a_{(0)} +  C^{a^c}_{(0)}\] {\rm Tr} F^2_a
\eea
where we sum over pairs $(a,a^c)$. The GS-diagram is proportional
\beq
N_a\(I_{ab} + I_{ab^c} - I_{a^cb} - I_{a^cb} \)
\eeq
which has the correct form to cancel the mixed chiral $U(1)_a SU(N_b)^2$ anomaly.

\subsection{Massless $U(1)$'s}

A nonzero coupling $C_{(2)} F$ generates an effective mass for the $U(1)$ gauge field. A linear combination
$$ \sum_i \theta_i F_i$$
is massless if and only if
\beq \label{eq:msslss}
\sum_i \theta_i N_i \[ C^i_{(2)} -  C^{i^c}_{(2)}\] = 0
\eeq 
This equation can have nontrivial solutions because the basis $a$ 
that we are using is highly non-degenerate. It is natural
 to expand this ``brane-basis" $\{C^{a}_{(2)}\}$ 
in a complete, non-degenerate basis $\{C^{(m,J)}_{(2)}\}$ 
that is one-to-one to the Ishibashi labels $(m,J)$
\beq \label{eq:basis}
C^a_{(2)} = \sum_{(m,J), w_m \neq 0} R_{a(m,J)} C^{(m,J)}_{(2)} \;\;\; .
\eeq
where $R$ are the boundary reflection coefficients. 
This equation has a well-known geometrical analogue, namely as an expansion of
the basis $\omega_a$ in a homology basis. In a geometric orientifold the tadpole conditions are
written in a convenient homology basis $\pi_i$ 
for $H^3(CY)$. The RR $2$-form in the brane basis can then be expanded as $C^a = \sum_{i} \pi_{ai} C^i$ where
$C_i$ are the reductions of the ten-dimensional RR$5$-form along the $\pi_i$ and $\pi_{ai}$ are the wrapping
numbers or charges  of the brane. From the CFT tadpole condition it follows that the 
Ishibashi labels are the natural basis at the Gepner point and that the reflection coefficients are the "wrapping
numbers". This is the motivation for~\R{basis}. Then~\R{msslss} becomes
\beq \label{eq:massless}
\sum_i \theta_i N_i \[R_{a(m,J)} - R_{a^c(m,J)} \] = 0
\eeq 
for all Ishibashi labels $(m,J)$.

\section{Results}\label{sec:results}
In this section we will present the results of our search. We list some
statistics and characteristics of the Gepner models and MIPFs we scanned in
tables  \ref{tbl:gepner_models}, \ref{tbl:mipfs_sols} and \ref{tbl:nr_sols}. 
Furthermore we display plots of distributions of standard model
particle multiplicities, the Higgs and some characteristics of the various
models like ratios of gauge couplings, the number of hidden branes and
features of the hidden gauge group. We will also present a small investigation
into varying the number of chiral families.

\subsection{The numbers}
In total we found \totSpectra\ distinct standard model spectra with solutions to
all tadpole equations. There are few cases where the same spectra are obtained 
for different MIPFs. In some cases these MIPFs have the same Hodge numbers and are
also otherwise indistinguishable. This occurs for example for the tensor product $(2,2,2,6,6)$
and presumably indicates an unresolved redundancy. In other cases, the Hodge numbers
are different, although, surpsisingly, the open sector is identical. 
The total number of such equivalent spectra is however very small, and if we remove these equivalences there
are 179119 left. In the rest of the paper we use the former set as the basis of our analysis.

In table \ref{tbl:gepner_models} we list for each Gepner model the search results.
The second column contains the values of the factors $k$ in the tensor product.
Models for which we only searched for solutions of type 0 and 1 are
denoted by a dagger $\dagger$. 
The third column specifies the number of primaries,
the fourth the number of simple currents and the fifth gives information
on the modular invariant partition functions. 
In this column, the first entry is the total number of symmetric simple 
current MIPFs.
This is computed after removing MIPFs that are related to each
other by permutations of the identical factors of the tensor product.
Generically, all MIPFs related to different simple current subgroups
and different matrices $X$ are distinct. However, in special cases
generically distinct simple current MIPFs coincide (the simples example
is $SU(2)$ level 2, where the generic A and D invariants coincide). In
 the table we list the number after removing such coincidences.
Between parentheses we indicate for how many of these MIPFs the tadpole
conditions were not completely solved for all standard model brane stack configurations;
however even in those cases we searched for all solutions with at most three 
extra branes (or at most two if the number of candidate branes was larger than 400).
The next entry in column four is the number of MIPFs for which at least one
standard model brane configuration was found, and the last entry in column 4 is the 
number of MIPFs for which a solution to the tadpole equations was obtained.
Column five gives the total number of standard model configurations, summed
over all MIPFs, column six gives the total number of configurations with
solutions to the tadpole conditions, and the last column gives the total
number of distinct (as defined in section \ref{sec:models_tp_alg}) standard model
spectra.

In total we found solutions to the tadpole equations for 44 of the 168 Gepner Models and
for \mipfsWsols\ of the \totMipfs\ MIPFs. For \mipfsWOinter\ MIPFs we did not even find
any standard model four stack configuration. In 649 cases there are four-stacks, but no 
solution to the tadpole equations and in 342 cases we could not rule out the existence of solutions.
The total number of standard model configurations that exist is \totInters; of these \intersWsols\ 
yield solutions. Many of the latter have the same standard model spectrum, which reduces the
total to \totSpectra.

The 168 models are listed in a
particular order, starting with the ones with the smallest number of factors. 
More or less coincidentally this order corresponds rather well to the degree of difficulty, in
decreasing order.
The number of (a,b,c,d) stacks is very small at the bottom
of the list, and increases to millions at the top, but only for particular 
tensor combinations. However, despite the large number of candidates, the 
tensor combinations at the top of the list yield very few solutions to the 
tadpole equations. 
Some tensor combinations with a large number of solutions have a recognizable 
feature in common: the values of $k+2$ typically takes values that factorize 
into powers of 2 and 3. However, there are counterexamples in both directions.

In table \ref{tbl:mipfs_sols} we list the \mipfsWsols\ modular invariants for 
which tadpole solutions were found. Column two specifies the MIPF in terms of 
the Hodge numbers of the corresponding type-IIB Calabi-Yau 
compactification\footnote{Hence $h_{11}$ is the
number of hyper multiplets (not including the universal one from the gravitational
sector) and $h_{21}$ the number of vector multiplets of the closed type IIB string before
orientifolding.}
and the
number 
of singlets in a Heterotic compactification. Unfortunately this does not always
specify the MIPF uniquely. However, in most cases they are distinguished by
the number of boundaries, listed in column three. Column 4 contains the 
(unique) label of the MIPF\rlap.\footnote{This label specifies a simple current
subgroup and a matrix $X$ that defines the modular invariant. It is difficult 
to list all this information efficiently, but it is available from the authors 
on request.}
In the last column we list the number of different Standard model spectra 
organized according to type. The first six entries refer to types 
$0,\ldots,5$ defined above, with a massless $B-L$ vector boson.
The last entry is the number of type 0 spectra with a massive $B-L$.
In principle this could have occurred for type 1 as well, but no examples were 
found.

Table \ref{tbl:nr_sols} lists the total number of solutions for each type, 
where we distinguish chiral subtypes for types 1,3, and 5. These subtypes
are defined in terms of the contribution of the quark doublets, the
lepton doublets and the Higgses to the $U(2)_{\rm b}$ anomaly. Since this 
anomaly
cancels, the subtypes are characterized by two independent parameters. It
is easy to see that the quark contribution must be an odd multiple of three
(which we have chosen to be positive), the lepton contribution must be odd,
and the Higgs contribution even. If the lepton contribution is larger than
three the spectrum contains ``chiral mirror leptons'', i.e. mirror pairs of
leptons that are chiral with respect to the full CP group, but non-chiral
with respect to the standard model gauge group. For example, to get a
lepton $SU(2)$ anomaly of $-5$ one must have the representation
$4 (1,2^*,-\frac12) + (1,2^*,\frac12)$ (up to purely non-chiral mirrors).
Strictly speaking such models can perhaps not be described as
``just the chiral standard model'', but we admitted them as a curiosity.
The same phenomenon is possible for quarks as well, but we did not find 
any examples. Note that the last column shows {\it twice} the number of chiral
supersymmetric Higgs pairs $(1,2,\frac12)+(1,2,-\frac12)$, so that the number of such Higgs pairs in the entire set can
be 0,1,2,3,4 or 6.
Interestingly only 0 and 3 were found for type 1.

\subsection{Features of found spectra}
As discussed above there are several features we have considered as relevant to 
distinguish standard model spectra. We will now present which values these 
parameters take and see if there is any notable structure. 

Looking at these distributions one could be tempted to draw statistical 
conclusions. Even if one would adopt this point of view, there are
several reasons why one should be careful. For one, our search algorithm was set
up to maximize the number of {\em different} solutions (see section \ref{sec:models_tp_alg}).
Also there are relations between different distributions (for example if stack 
c is realized as $SO(2)_{\rm c}$, $e^c$ and $\nu^c$ come from the same branes, hence
have the same mirror-distribution).
Another bias to keep in mind is caused by the fact that equations are much
easier to solve for a small number of branes. This biases the search towards fewer branes.

First we will look at the number of mirrors of the standard model particles. 
The total chirality of the standard model particles is fixed to 3. We allowed 
however for additional non-chiral pairs, the so-called mirrors.
The plot for the number of mirrors is similar for all standard model particle
hence we show only the one $(e,\nu)$ mirror pairs and the total number of mirrors
of standard model particles in a model.
\begin{figure}[!ht]\label{fig:mirrors}
\caption{Mirrors}
\includegraphics{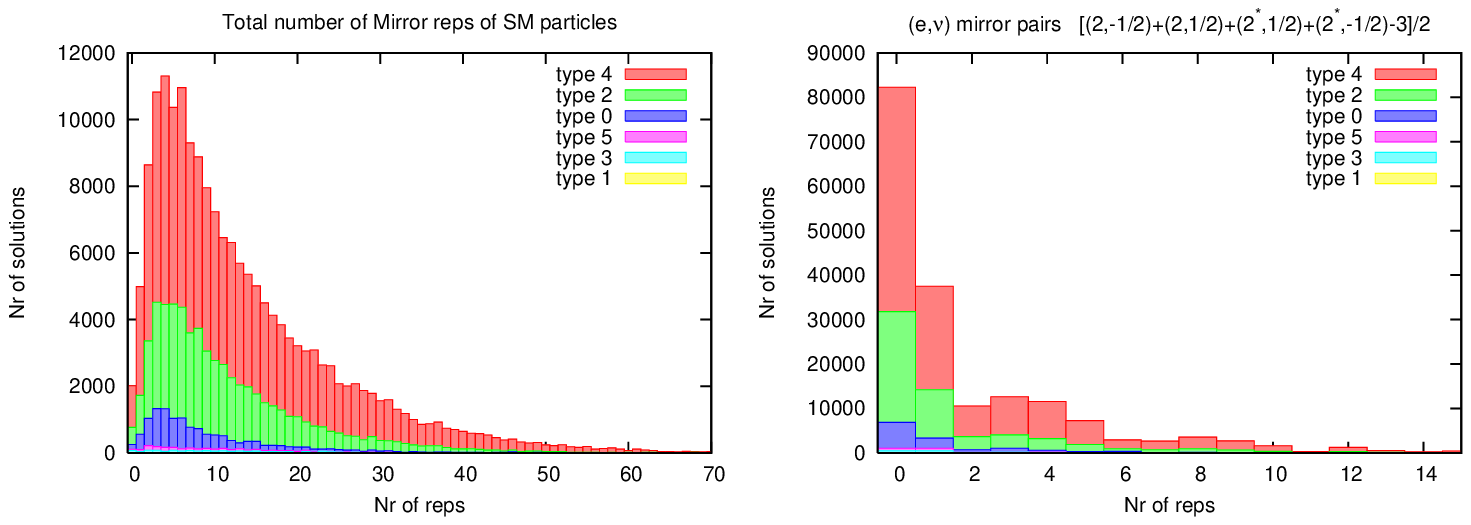}
\end{figure}
It is interesting to see that the distribution of number of mirrors is 
sharply peeked at zero mirrors. 
From the total plot we see that \noMirrors\ models have no mirrors at all and that the
distribution peeks at 4 mirrors.

\begin{figure}[!ht]\label{fig:leptoquarks}
\caption{Non-Chiral Leptoquarks}
\includegraphics{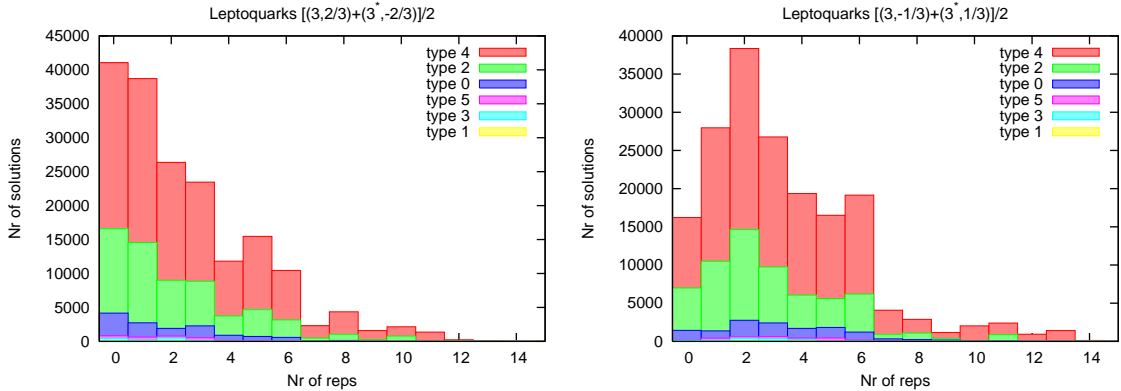}
\end{figure}

The only bifundamentals coming from string states between the standard model
branes which should not be chiral are strings stretched between branes a and
d. These particles would be leptoquarks. In figure \ref{fig:leptoquarks}
we plot the number of non-chiral leptoquarks.
The distribution of leptoquarks with equal sign lepton and baryon number
is peaked at zero. This is the only non-chiral 'exotic' that peaks at zero (apart from
individual mirror distributions, as noted above).
The distribution of the opposite sign leptoquarks distribution peaks at 2.

\begin{figure}[!ht]\label{fig:tensors}
\caption{Non-Chiral tensor representations}
\includegraphics{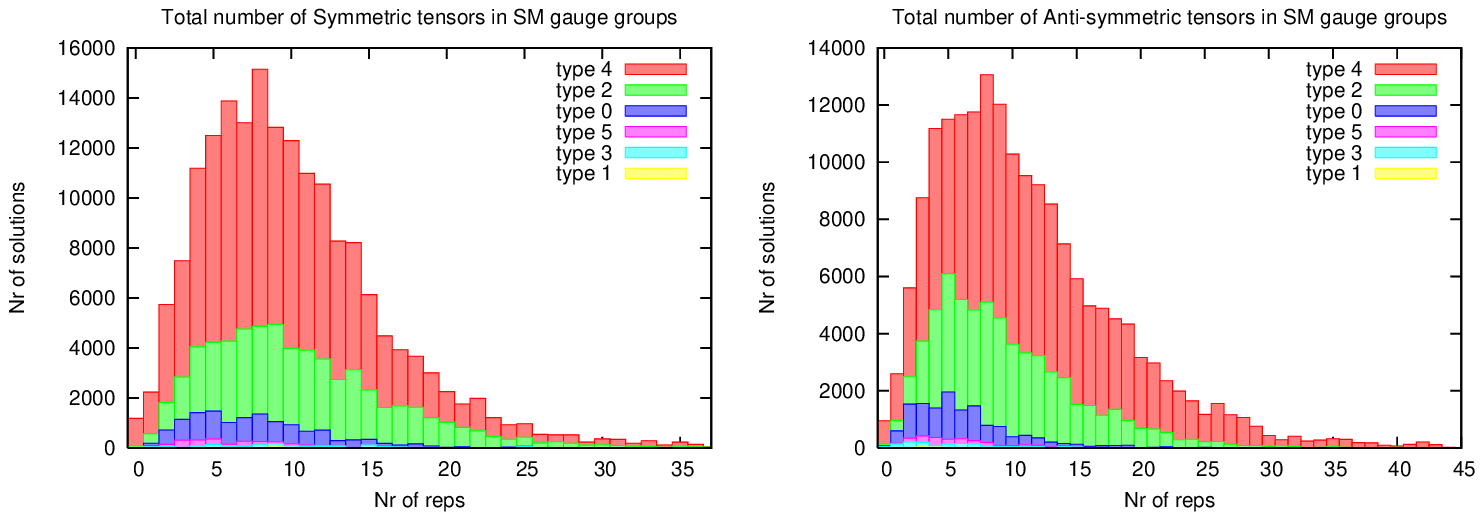}
\includegraphics{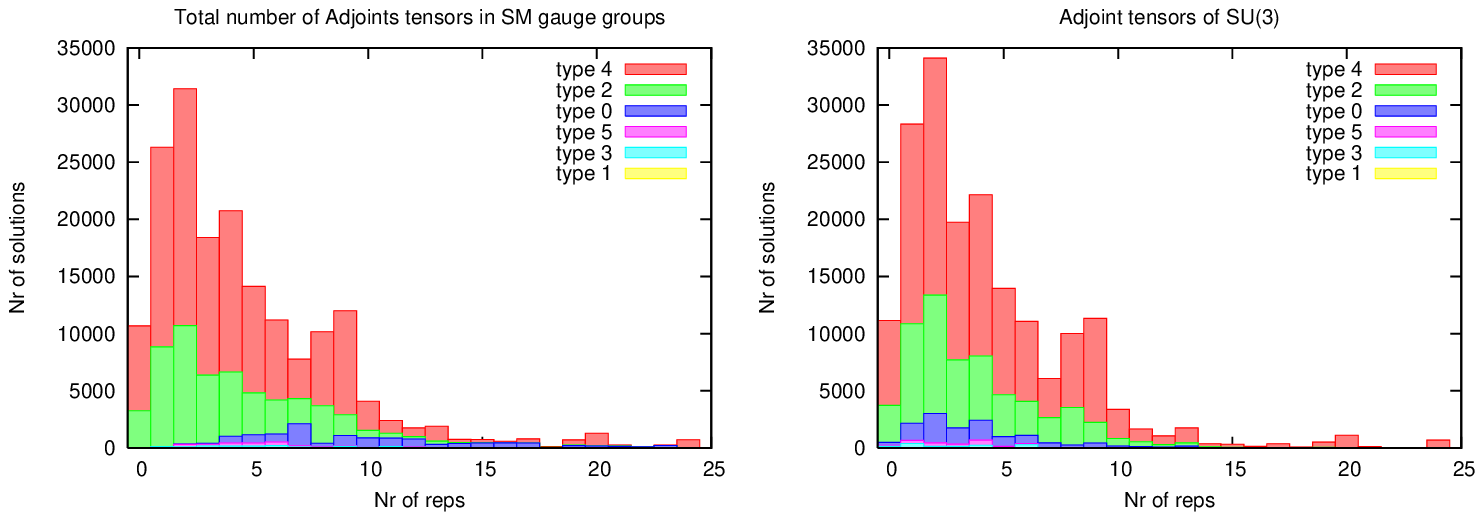}
\end{figure}

Finally one can have massless non-chiral representations coming from strings that
have both ends on the same SM brane (or one end on the conjugate brane). These
states are symmetric, anti-symmetric or adjoint tensors of the standard model 
gauge group. The distributions of non-chiral pairs of all these particles
peak at a non-zero value, except for the 
ones where in a majority of the models that particular state does not 
exist\rlap.\footnote{Adjoints of $SO(2)$ are counted as anti-symmetric tensors,
adjoints of $Sp(2)$ as symmetric tensors. Massless anti-symmetric representations of
$U(1)$ actually have no massless states at all, and were not counted.}
In \ref{fig:tensors} we plot the total number of these particles in 
a model and the distribution of adjoints of $SU(3)$ as an example. 

\subsection{Higgs}
\begin{figure}[!ht]\label{fig:higgs}
\caption{Number of Higgs}
\includegraphics{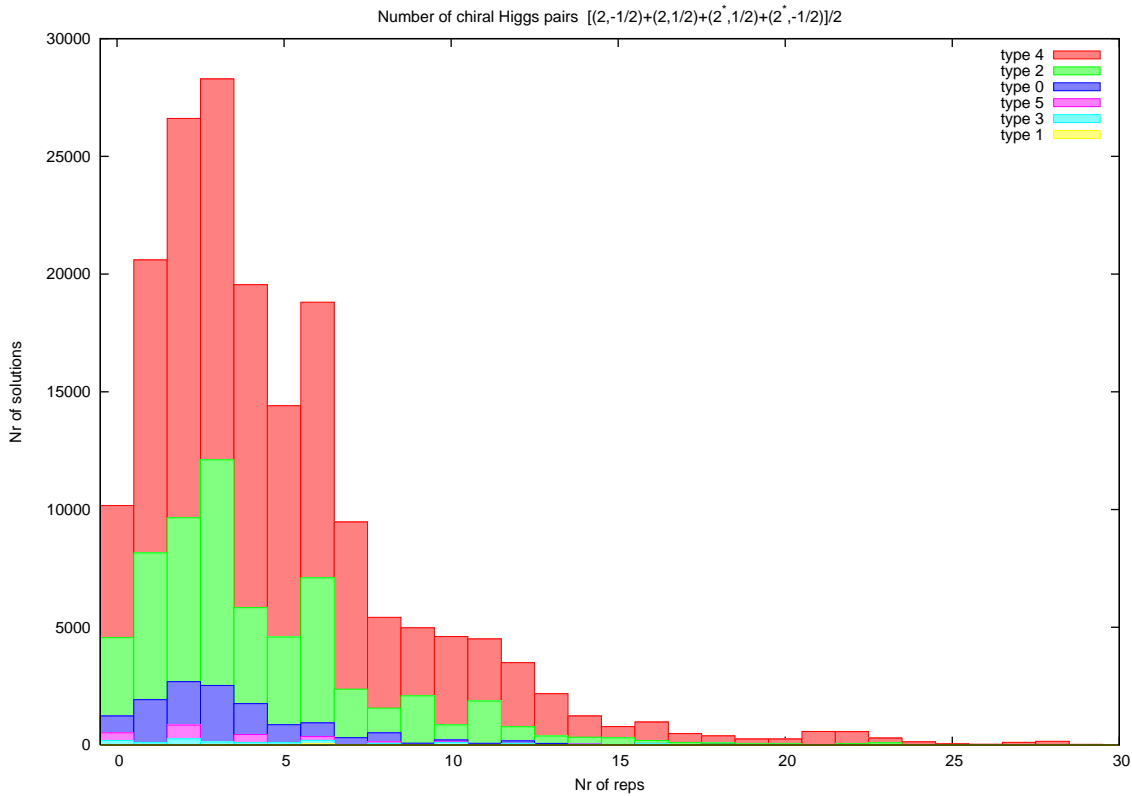}
\end{figure}

In supersymmetric extensions of the standard model the Higgs always comes
in non-chiral (w.r.t. to $G_{SM}$) pairs so the Higgsinos will not generate an
Y-anomaly.
In figure \ref{fig:higgs} we plot the number of Higgsino pairs,
which is thus equal to half the number of standard model Higgs doublets in that model.
As is clear from the picture, the number of Higgs pairs peaks at three. The 
maximum number of Higgs pairs we found is \maxHiggs. Note that there also models
with no Higgs. These models have an obvious deficiency, although it is
conceivable that some (composite-)particle from the hidden sector will play
the r\^ole of the Higgs.

For types based on $U(2)_{\rm b}$ (type 1,3 and 5) there is a possibility for
the Higgs of being chiral with respect to $U(2)_{\rm b}$. This chirality was ignored
in figure \ref{fig:higgs}, but the chirality distribution is display in \ref{tbl:nr_sols}, as
explained above.

\subsection{Hidden branes}
\begin{figure}[!ht]\label{fig:numberofbranes}
\caption{Hidden sector of all solutions}
\includegraphics{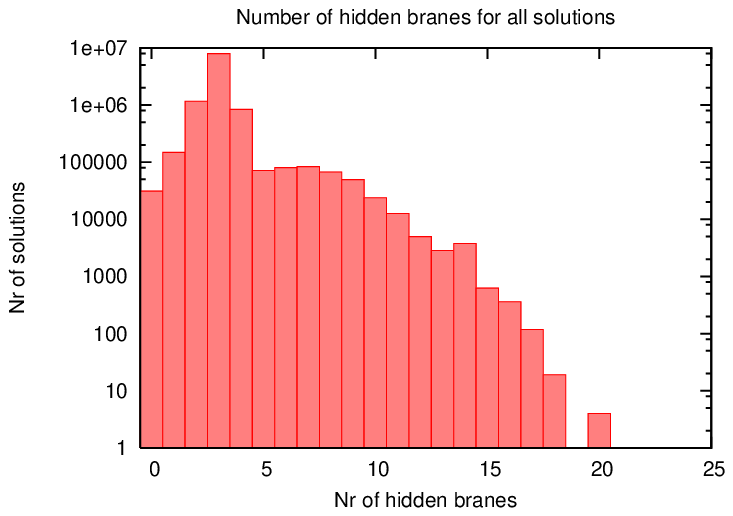}
\includegraphics{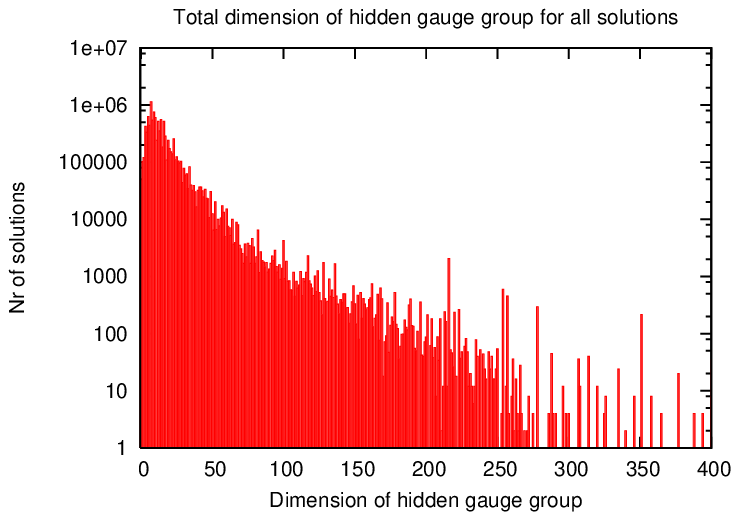}\label{fig:ggdim}
\end{figure}

Our search algorithm was set up to maximize the different standard model
spectra, not the hidden brane degrees of freedom. Hence we present in all
plots and tables the number of solutions where we identified solutions 
with different hidden sectors, but which are otherwise equivalent.
As explained in section \ref{sec:models_tp_alg} we did however 
construct all solutions containing 0,1 and 2 extra branes for all standard model brane stacks, three extra branes for all
stacks with less than 400 candidate hidden CP groups, and four extra
branes for less than 100 candidate hidden CP groups.  
In many other cases we attempted to extract a
solution from the tadpole equations without a limit on the number of branes. The latter searches were
stopped as soon a one solution was found and are limited
by computer time constraints.   Therefore they are not systematic.
In figure \ref{fig:numberofbranes}
we show the total number of solutions found for each hidden brane multiplicity.
This plot is based on a total of 10526078 solutions.
These solutions are different only in the
sense that their CP multiplicities are distinct. Undoubtedly there are still some
equivalences in this set that were not taken into account. The number of solutions
with 0,1 and 2 branes is 31215, 148324 and 1170556 respectively.
As one can see, the number of solutions grows very fast
with the number of hidden branes, but is cut off beyond three branes for the
reasons explained above. One can make
a plausible guess what the picture might really look like, had we been able to push the 
search for solutions much further. Most likely, it would continue growing dramatically
for quite a while. Since the number of candidate CP groups is typically a few hundred, the
distribution necessarily peaks well below that number, but this could easily be around ten or
twenty. It is clear that the total number of solutions with distinct standard model plus hidden
spectrum can be many orders of magnitude larger than those we found.
In figure \ref{fig:ggdim} we plot the total dimension of the total hidden gauge 
group. 
The biggest gauge group we encountered has dim \maxdim. 
The biggest factors we found are $SO(\maxSO)$, $Sp(\maxSP)$ and $U(\maxU)$.

\subsection{Gauge couplings} 

A remarkable property of the standard model is that its gauge group fits 
naturally within the gauge group $SU(5)$. This explains two empirical facts: 
the observed quantization of electric charge, and the observed convergence of the 
coupling constants at short distances (although with present data, the latter 
success only survives if the additional assumption of supersymmetry is made). 
Both of these nice features seem to be lost in the class of models considered 
here.

Fractional charges are inevitable if there exist additional branes outside the 
standard model. Strings stretching between the standard model and the hidden 
branes yield particles (fundamental or QCD bound states) with half-integer electric charge. Although they may be
non-chiral (as they are in all models presented here) or even completely absent 
from the massless spectrum, they exist inevitably as massive open string states.
Interestingly, it is not completely straightforward to realize group-theoretical 
unification in heterotic strings either.
In the standard realizations with Kac-Moody level 1 one ether gets $SU(5)$ 
without a massless Higgs boson to break it, or one gets
$SU(3) \times SU(2) \times U(1)$ with additional (though not necessarily 
massless) fractional charges \cite{Schellekens:1989qb}. These problems can be avoided, for example by
considering higher Kac-Moody levels, but it is difficult to argue that charge
quantization is a natural property of string theory.

Heterotic $SU(5)$ and $SO(10)$ models do explain the observed coupling constant 
convergence, but they make a troublesome prediction for the unification scale, 
which is off by two to three orders of magnitude. It is on this point that open string
models have an advantage, simply because there is no such prediction \cite{Witten:1996mz}. But there
is also no prediction for the unification of the gauge couplings, because they 
emerge from dilaton couplings of four, {\it a priori} unrelated, branes. 

In \cite{Blumenhagen:2003jy} it is argued that a realization of a 
supersymmetric extension of
the standard model with intersecting branes naturally leads to a model
where respectively branes $a$ and $d$  and $c$ and $b$ wrap the same cycles.
This leads then at the string scale to the following relation between the three 
standard model coupling constants:
\begin{equation}\label{eq:bls_line}
\frac{1}{\alpha_Y}=\frac{2}{3}\frac{1}{\alpha_s}+\frac{1}{\alpha_w}.
\end{equation}
That would mean that these models do not necessarily have full unification,
but they do reproduce a relation which is compatible with the $SU(5)$ relation
between the coupling constants.

To do a proper calculation of gauge unification we would have to assume a 
unification scale and evolve the couplings downward from that scale. Since we
are in a rational point of the moduli space, this scale is fixed to a value of order 
the string scale. The massless spectrum we obtain contains, in general,
several non-chiral particles that are presumably artifacts of the rational
points. If one assumes that all these particles remain massless until the TeV
scale, one should take them into account in the renormalization group flow. More 
ambitiously, one could work out the masses and the gauge couplings as a function
of the moduli near the rational point, and take all of this into account. Here we
will only address a simpler question, namely if there is any evidence for 
relations among the couplings at the string scale. If there is no relation, 
then clearly one might still get the correct low energy gauge couplings by 
starting outside the unification point in the space of couplings, and compensate 
this with exotic matter in the renormalization group flow, but the concept of 
unification is then anyway lost.

The coupling constants can be computed as follows.
Up to a universal factor , they are given by
\beq
\frac{1}{g_{\rm a}^2} = R_{0{\rm a}} e^{\phi}
\eeq
for CP-groups of orthogonal and symplectic type and
\beq
\frac{1}{g_{\rm a}^2} = (R_{0{\rm a}}+R_{0a^c}) e^{\phi}  = 2 R_{0{\rm a}} e^{\phi}
\eeq
for unitary groups, with the conventional normalization $\rm{Tr~} T^aT^b=\frac12 \delta^{ab}$
for the generators in all three cases. This immediately gives the following expression for
the ratio of weak and string couplings:   
\beq
{g_2^2 \over g_3^2} = {R_{0{\rm a}} \over \kappa R_{0{\rm b}}}
\eeq
where $\kappa=1$ for spectra of types 1,3 and 5, and $\kappa=\frac12$ for spectra of type 0,2 and 4. 
The canonically normalized $U(1)$ generator of stack a is $Y_{\rm a}=\frac{1}{\sqrt{6}}\bf{1}$, for stack d it is  
$Y_{\rm d}=\frac{1}{\sqrt{2}}\bf{1}$,
and for stack c it is $Y_{\rm c}=\frac{1}{\sqrt{2}}\bf{1}$ for complex boundaries, and $Y_{\rm c}=\frac12 \sigma_3$ for real ones.
 The
standard model $U(1)$ factor $Y$ is the given by 
\beq
{1\over \sqrt6 } Y_{\rm a} - {1\over \sqrt2 } Y_{\rm c} -  {1\over \sqrt2 } Y_{\rm d}
\eeq
for types 0 and 1, and           
\beq
{1\over \sqrt6 } Y_{\rm a} -  Y_{\rm c} -  {1\over \sqrt2 } Y_{\rm d}
\eeq
for types 2,3,4 and 5.
This leads to the following expression for the coupling constants
\beq
 {1\over g_Y^2} = \frac16 {1\over g_{\rm a}^2}+\frac12 {1\over g_{\rm c}^2} +\frac12 {1\over g_{\rm d}^2}
\eeq
for types 0 and 1, and   
\beq
 {1\over g_Y^2} = \frac16 {1\over g_{\rm a}^2}+{1\over g_{\rm c}^2} +\frac12 {1\over g_{\rm d}^2}
\eeq
for types 2,3,4 and 5. 
Hence in both cases the expression for the couplings in terms of reflection coefficients is the same:
\beq
 {1\over g_Y^2} = ({1\over 3} R_{0{\rm a}}+R_{0{\rm c}} +R_{0{\rm d}})
\eeq
The fact that the same expression is obtained is due to the fact that an 
orthogonal or symplectic c-stack can be viewed as a limit of a unitary one, with the two conjugate branes moved
on top of the orientifold plane.  From this expression we obtain the
following formula for ${\rm{sin}}^2 \theta_W = g_Y^2/(g_2^2+g_Y^2)$:
\begin{equation}
\sin^2 \theta_W = \frac{\kappa R_{0{\rm b}}}{\kappa R_{0{\rm b}} + \frac16 R_{0{\rm a}}+\frac12 R_{0{\rm c}} +\frac12 R_{0{\rm d}}}
\end{equation}

\begin{figure}[!ht]\label{fig:couplings}
\caption{Gauge coupling constants at the string scale}
\includegraphics{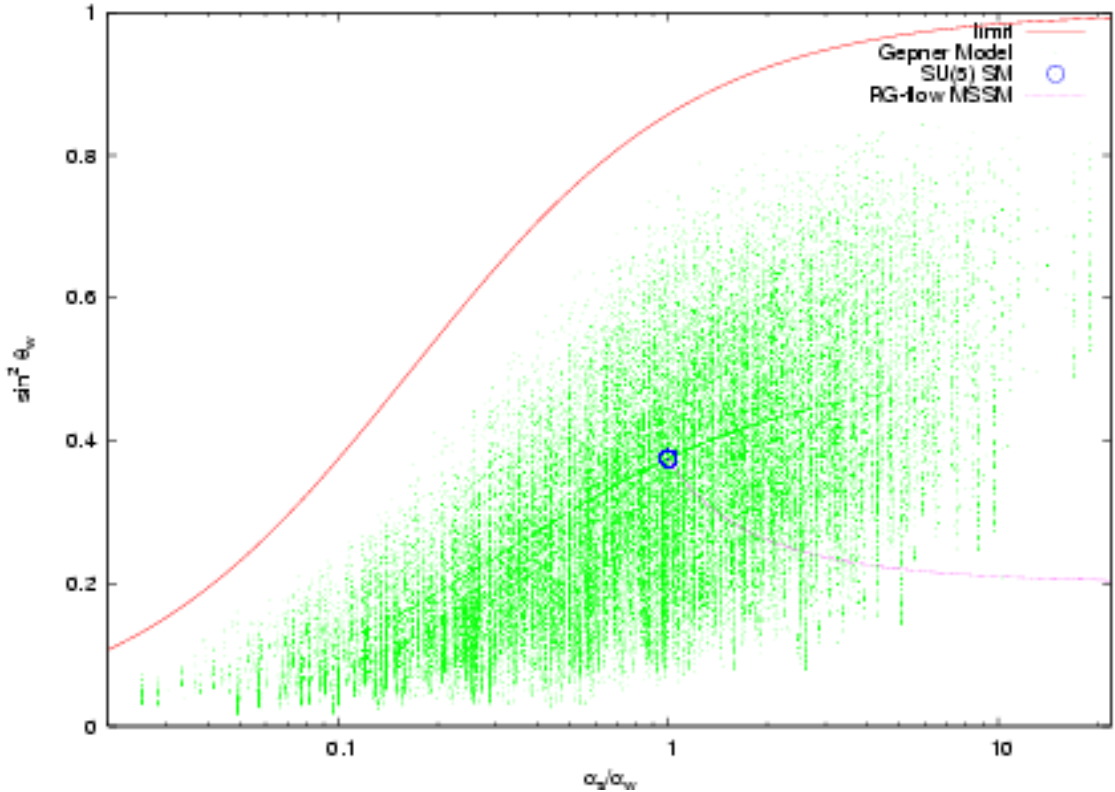}
\end{figure}

In figure \ref{fig:couplings} we plot $\sin^2(\theta_w)$ against the ratio
$\alpha_s/\alpha_w$. In this plot the value for $SU(5)$ unification
are also indicated, as well as its renormalization group flow.  The solid line
denotes the upper limit on $\sin^2(\theta_w)$. Clearly the models we found
occupy a substantial part of the allowed region.
It is amusing to see that this point falls neatly in the
middle of the cloud formed by our models.
One aspect of the plot that attracts attention is a faint line \footnote{This line is not very visible in the picture. A picture with higher resolution can be downloaded from http://www.nikhef.nl/$\sim$t58/GaugeCouplings.ps} roughly in the 
middle of the cloud. It turns out to be described by 
relation (\ref{eq:bls_line}). Approximately 10\% of the models is on that line.
Not surprisingly, these models are characterized by the fact that the 
reflection coefficients for respectively branes $a$ and $d$ and branes $b$ and 
$c$ are the same, which 
is precisely the RCFT equivalent of the condition that 
\cite{Blumenhagen:2003jy} used for unification.
So although a portion of the models we found has a
relation between the coupling constants compatible with $SU(5)$ unification,
this is certainly not a generic feature.

\subsection{Varying number of chiral families}

\begin{figure}[!ht]\label{fig:var_fam}
\caption{Solutions for varying number of chiral families}
\includegraphics{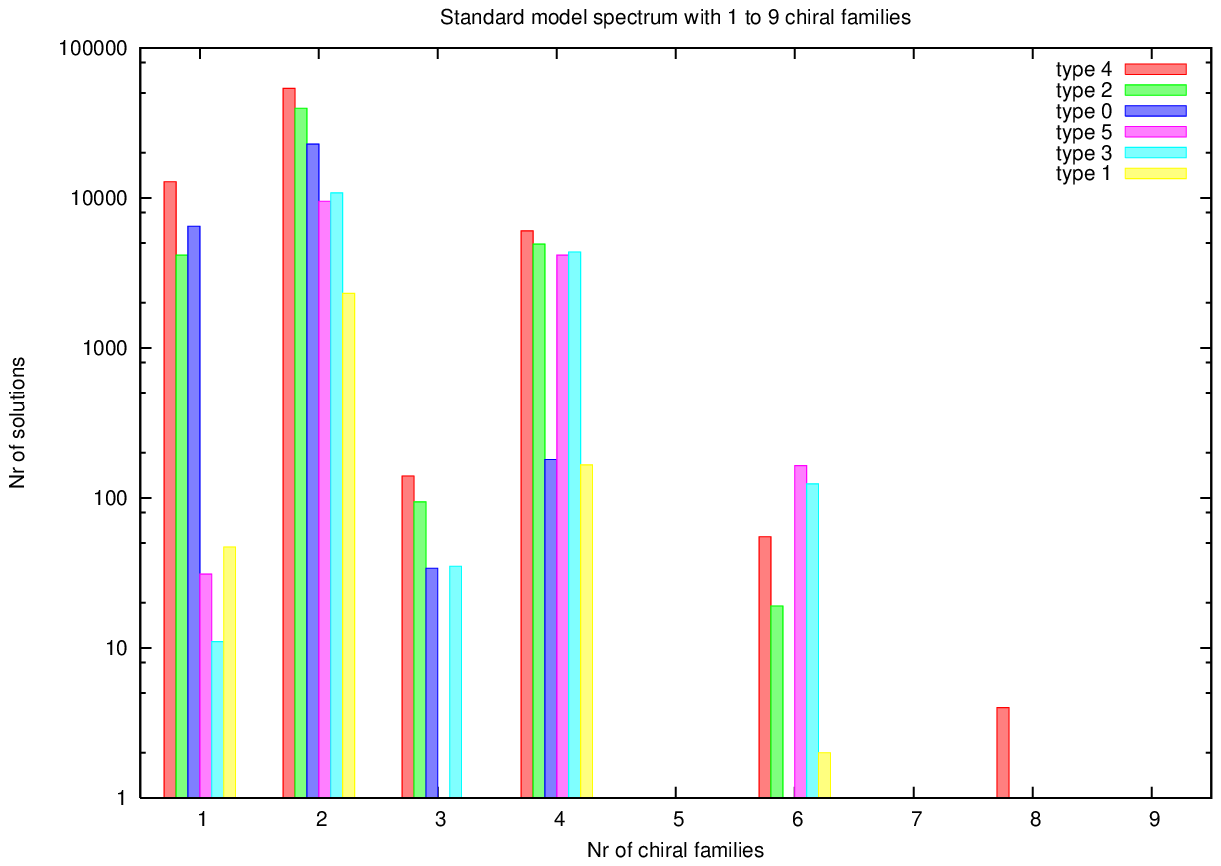}
\includegraphics{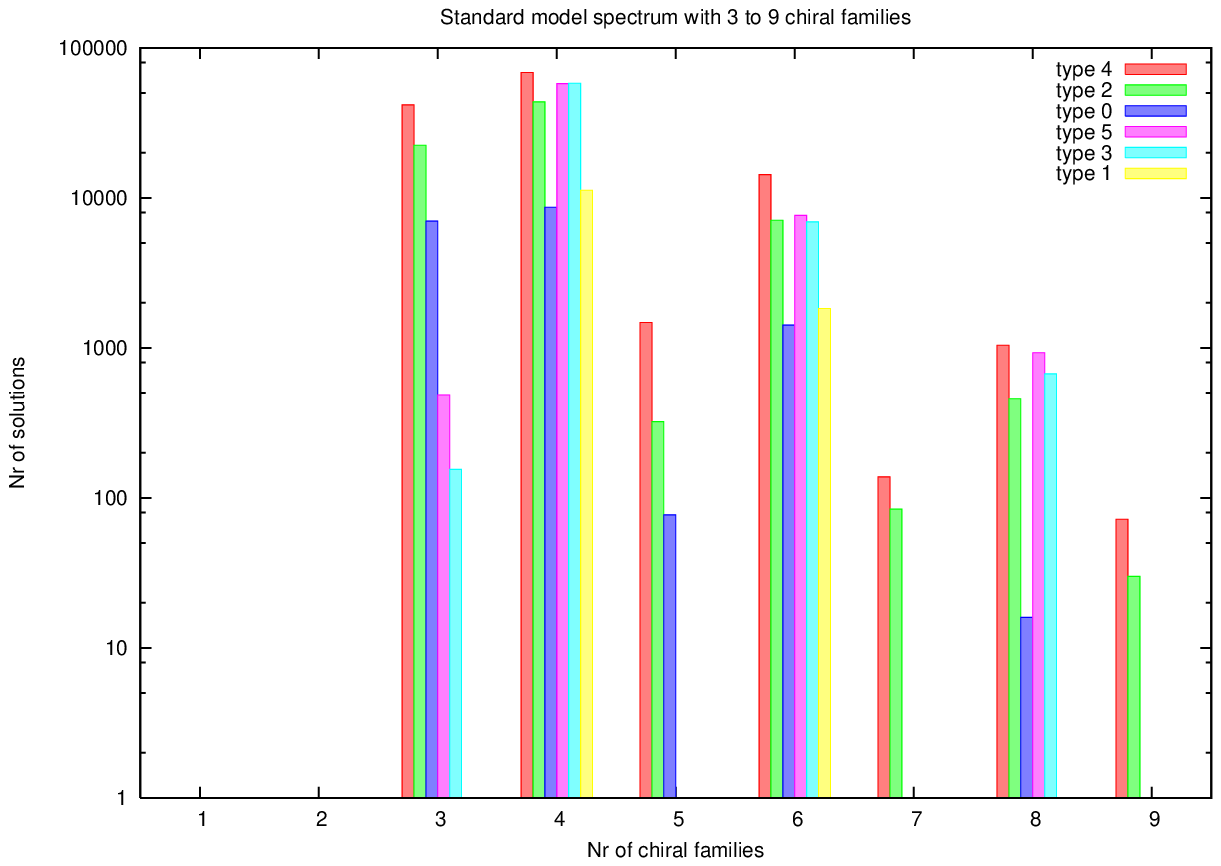}
\end{figure}

For a limited set of models we repeated our search for the standard model
spectrum with a different number of families, in order to get a feeling what the
constraint of having 3 chiral families means.
Because more chiral particles implies more intersections one could expect
that getting a spectrum with more chiral families is harder. Indeed this
tendency is visible in plots \ref{fig:var_fam}.
Because the number of standard model brane configurations to be analyzed
for 1 and 2 families was so big, we performed a search
for 1 up to 9 families for \mipfsFaMonetonine~models and a search for 3 up to 9 
models for a bigger set of \mipfsFaMthreetonine~models. 

But there is more structure. For example odd types (the ones based on 
$U(2)_{\rm b}$)
are much more abundant with {\it even} families. If one takes a closer look 
and distinguishes by subtype, that is the way the 
$U(2)_{\rm b}$ anomaly cancels, this becomes a bit clearer.
Both states coming from $[b,x]$ and $[b^*,x]$ can contribute to the total 
chirality of the $SU(2)_{W}$ doublets.
They do however contribute oppositely to the $U(2)$ anomaly. 
This leaves the possibility to cancel the anomaly between the different $U(2)$ chiralities
of a doublet.
In the even family number models solutions of the subtype where the anomaly 
cancels fully between the different $U(2)$ chiralities of a given doublet 
are approximately three orders of magnitude more abundant than subtypes
where the anomaly cancels partly between different doublets.
Apparently it is easy to have the intersection number of $[b,x]$ equal to 
that of $[b^*,x]$.
This is also explains why there are so relatively few odd family number
solutions with $U(2)_{\rm b}$; For the family number to add up to an odd number,
one of the two intersection numbers $[b,x]$ and $[b^*,x]$ has to be odd 
and the other even. The {\em easy} way of cancelling the anomaly is not a 
possibility.

Under the assumption that more chiral intersections are harder to get, 
we can also 
understand why 4, 2, 3 and 5 are typically less abundant
then types 0 and 1. The latter types are based on $U(1)_{\rm c}$ and have 
all weak singlets coming from their own intersection, with $c$ or $c^c$.
The former types however,
have a real $c$-brane and both lepton singlets (and both quark singlets) 
come from the same intersection.
This means that they have 2 times the number of families less chiral 
intersections. 

A similar effect doesn't occur if we exchange $U(2)_{\rm b}$ for 
$Sp(2)_{\rm b}$, because as discussed above, both $U(2)_{\rm b}$ chiralities contribute to one doublet. Hence both types based on $U(2)_{\rm b}$ and 
$Sp(2)_{\rm b}$ have the same number of chiral $b$-brane intersections.

Also in the other types there is a clear even/odd effect. But if one
looks, for odd/even family number separately, at the trend in the number
of solutions for a certain type against family number it is to good 
approximation an exponential drop.

One feature that deserves mentioning is the occurrence of type 1 models
with a massive $B-L$ for 1,2 and 4 families, something we didn't find 
for three families. These models have as the gauge group that couples to
the standard model particles precisely the standard model. That this didn't
occur for three families is probably a matter of statistics. As we have seen
type 1 models with three families are already rare, and furthermore from the
search for 1,2 and 4 family number models we know that a massive $B-L$ only 
happens in a few percent of the type 1 solutions.

\section{The simplest case}\label{sec:simplest}

Here we will discuss in some detail one of the simplest cases where
a solution was found. It occurs for the tensor product $(1,1,1,1,7,16)$.
This has 1944 primaries, 81 of which are simple currents forming a
discrete group $\ZZ_3 \times \ZZ_3 \times \ZZ_3 \times \ZZ_3$. Since there are
only odd factors, there is just one orientifold choice for each MIPF.

After removing all permutations of the identical factors, this tensor
product has 34 simple currents MIPFs. Only one of them has standard-model-like
(a,b,c,d) branes. This MIPF is defined by a simple current group that is
isomorphic to $\ZZ_3 \times \ZZ_3 \times \ZZ_3$, and is generated by the three
currents $[0,0,0,0,(1,-1,0),(7,3,0),0]$,
$[0,0,0,0,(1,-1,0),(7,3,0),0]$ and $[0,0,(1,-1,0),0,0,(7,-5,0),(16,16,0)]$. Out notation
is as follows: between the square brackets we indicate the decomposition
of the tensor product representation into minimal model representations, with the
first entry in each square bracket representing the space-time sector. The minimal
model representations are indicated in the usual way as $(l,q,s)$, except for the vacuum,
which is denoted as "0". Note that only orbit representatives are shown: The minimal
model representations are subject to field identification, and the entire tensor
representation is actually just one representative of a chiral algebra orbit. Hence
other, but equivalent, expressions can be given.

The matrix $X$ yielding the MIPF under consideration is
\begin{equation*}
\frac13\begin{pmatrix}
 1 & 0 & 1 \\
 0 & 1 & 1 \\
 1 & 1 & 1  \\
 \end{pmatrix}
\end{equation*}

This yields a closed string spectrum corresponding to Hodge numbers $(23,23)$, and
a heterotic spectrum with 217 singlets.
There are 72 Ishibashi states and hence the same number of boundary states. The total
number of sets of (a,b,c,d) stacks is six, all of them of type 3. Three sets saturate the dilaton tadpole
condition, whereas the other three do not. We find that for the former all other 17
tadpole conditions are also satisfied, whereas for the latter three there is no
possibility of adding branes in order to solve the tadpole conditions. The
three sets of boundaries yielding solutions can be described as follows
\begin{align*}
\text{set 1}:\ \  &({\rm a}_1,{\rm a}_1^c)({\rm b},{\rm b}^c)({\rm c})({\rm {\rm d}},{\rm d}^c) \\
\text{set 2}:\ \  &({\rm a}_2,{\rm a}_2^c)({\rm b},{\rm b}^c)({\rm c})({\rm {\rm d}},{\rm d}^c) \\
\text{set 3}:\ \  &({\rm a}_3,{\rm a}_3^c)({\rm b},{\rm b}^c)({\rm c})({\rm {\rm d}},{\rm d}^c) \\
\end{align*}
in other words, the b,c and d branes are always identical, and the
complete set of boundaries we need is given by
\begin{align*}
{\rm a}_1:& [0,0,0,0,0,(1,-1,0),(2,2,0)]  \\
{\rm a}_1^c:& [0,0,0,0,0,(1,1,0),(2,-2,0)]  \\
{\rm a}_2:& [0,0,0,0,0,(1,-1,0),(8,2,0)]_1  \\
{\rm a}_2^c:& [0,0,0,0,0,(6,4,0),(8,-8,0)]_1  \\
{\rm a}_3^c:& [0,0,0,0,0,(1,-1,0),(8,2,0)]_2  \\
{\rm a}_3^c:& [0,0,0,0,0,(6,4,0),(8,-8,0)]_2  \\
{\rm b}_{\phantom 1}:& [0,(1,-1,0),(1,-1,0),(1,-1,0),(1,-1,0),(6,-6,0),0]  \\
{\rm b}^c:& [0,(1,1,0),(1,1,0),(1,1,0),(1,1,0),60,0]  \\
{\rm c}_{\phantom 1}:& [0,0,0,0,(1,1,0),(7,7,0),(16,16,0)]  \\
{\rm d}_{\phantom 1}:& [0,0,0,0,0,(7,7,0),(16,-14,0)]  \\
{\rm d}^c:& [0,0,0,0,0,(7,-7,0),(16,14,0)]  \\
\end{align*}
The three sets only differ in the choice of the $U(3)$-brane. The last two
choices, ${\rm a}_2$ and ${\rm a}_3$ and their boundary conjugates,
are resolved fixed points of the supercurrent extension of the
tensor product., and they differ only in their
fixed point degeneracy labels, indicated here as 1 and 2.
Not surprisingly, they yield the same spectrum.
The spectra of sets 1 and 2 are distinct, and are as follows
\begin{eqnarray*}
     &5  ( V ,V^*,0 ,0 )_3     +
      9  ( V ,0 ,V ,0 )_{-3}  +\\
      &5  ( 0 ,V ,V ,0 )_3     +
      3  ( 0 ,V ,0 ,V )_3     + \\
      &3  ( 0 ,0 ,V ,V )_{-3}   +
     12  ( V ,V ,0 ,0 )_0      +\\
      &4  ( 0 ,V ,0 ,V^*)_0     +
      8  ( Ad,0 ,0 ,0 )_0      +\\
      &4  ( A ,0 ,0 ,0 )_0     +
      8  ( S ,0 ,0 ,0 )_0     +\\
      &4  ( 0 ,A ,0 ,0 )_0     +
      3  ( 0 ,0 ,S ,0 )_0     +\\
      &4  ( 0 ,0 ,0 ,A )_0     \\
\end{eqnarray*}
and
\begin{eqnarray*}
      &3  ( V ,V^*,0 ,0 )_3    +
      9  ( V ,0 ,V ,0 )_{-3}  +\\
      &5  ( 0 ,V ,V ,0 )_3      +
      3  ( 0 ,V ,0 ,V )_3      +\\
      &3  ( 0 ,0 ,V ,V )_{-3}   +
      4  ( V ,V ,0 ,0 )_0      +\\
      &4  ( 0 ,V ,0 ,V^*)_0     +
      4  ( Ad,0 ,0 ,0 )_0    +\\
      &4  ( A ,0 ,0 ,0 )_0    +
      6  ( S ,0 ,0 ,0 )_0    +\\
      &4  ( 0 ,A ,0 ,0 )_0    +
      3  ( 0 ,0 ,S ,0 )_0    +\\
      &4  ( 0 ,0 ,0 ,A )_0   \\
\end{eqnarray*}
The notation is as follows: $V$ denotes a vector, $A$ and anti-symmetric tensor, $S$ a
symmetric tensor and $Ad$ and adjoint representation. The $*$ denote complex conjugation,
and the last subscript denotes the net chirality of the representation. Our convention
is to represent all four-dimensional fermions as left-handed spinors. The precise meaning
of, for example, $N \times (V,V^*,0,0)_M$ is then $\frac12{(N+M)} (V,V^*,0,0) + \frac12{(N-M)}(V^*,V,0,0)$.
Hence in the first case, we have 3 chiral standard model
$SU(3) \times SU(2) \times U(1)$ multiplets $(3,2,\frac16)$, plus a single
mirror pair $(3,2,\frac16)$+$(3^*,2,-\frac16)$. Note that the second factor is
actually $U(2)$ instead of $SU(2)$. The additional $U(1)$ is anomalous, and the
corresponding gauge boson acquires a mass. The mirrors are non-chiral not just
with respect to the standard model group, but with respect to the full group (although
this was not {\it a priori} required). There is a second set of fully non-chiral mirrors
of the same standard model representation in both cases, six in the first and 2 in the
second. These differ from the former ones by having opposite $U(1)_{\rm b}$-charges.
The tadpole equations satisfied by these two cases are in fact identical.

Since these spectra only differ by the choice of the $a$-branes, the lepton and Higgs
sectors are identical. Both have 5 Higgs pairs $(1,2,\frac12)+(1,2,-\frac12)$, with
the interesting feature that 3 of them are chiral with respect to the $U(2)$ group.
Another noteworthy feature is the absence of lepto-quarks.

Finally we present the closed spectrum.
The only model-dependent feature concerns the diagonal sector. It turns out
that all relevant multiplicities are 1, and that all non-vanishing Klein bottle coefficients
are 1. The 23 closed string vector multiplets originate from 21 diagonal fields in the closed string partition function,
and two off-diagonal ones. Of the first 21 only the symmetric components survive the
Klein bottle projection, and the two off-diagonal ones are reduced to a single field,
consisting of a symmetric and an anti-symmetric component. Hence the projected
closed string spectrum consists of a supergravity multiplet,
one vector multiplet and 22 chiral multiplets originating from $h_{11}$ and another 23 chiral multiplets
originating from $h_{21}$.

\section{Conclusions}\label{sec:conclusions}

We performed the first semi-systematic search for standard model spectra in 
open string models. We presented \totSpectra~models that have as their chiral 
spectrum just the standard model. Most of the models do have non-chiral exotics 
and/or mirrors, for which we presented the abundances. The number of Higgs was
left a free parameter. We found models with as much as \maxHiggs~Higgses. The
distribution peaks at
three pairs of Higgs. We also calculated  the standard model
gauge couplings and found (perhaps not surprisingly) no relation that
hinted at unification.

Although we are confident that within the Gepner models we
constructed most realizations of the ``standard'' four stack model, there is
much room for improvement and generalization.

More MIPFs of the Gepner models are known than the ones considered here. First of all
there are diagonal invariants and their simple current modifications. It turns out that in
many cases (but not all) these can be obtained by simple currents, and hence are already
included. In other cases more work is needed, see {\it e.g.} \cite{Birke:1999ik}.
Furthermore there are exceptional $SU(2)$ invariants, exceptional invariants of $SU(2)$
tensor products \cite{Schellekens:1990jh}, and invariants related to interchanges of fixed points.
Unfortunately there is no boundary state formalism available to deal with these cases.
There also
is no guarantee that we found all orientifolds. Up to now, the orientifold degrees of freedom we use here
(Klein bottle currents and crosscap signs) has included everything known by other methods, but most likely
this is just a conjecture awaiting a counter example.

There are additional $N=2$ coset CFT´s and presumably even RCFT´s that are not cosets.
Indeed,
there is a the much larger class of Kazami-Suzuki \cite{Kazama:1988qp} models. In principle
our methods are applicable to these models as well, provided we know the exact
spectrum. In cases without field identification fixed points this can be computed
straightforwardly, but if there are fixed points the formalism has not been
developed yet. Another class that might be accessible with similar methods are
the interchange orbifolds of identical minimal or Kazami-Suzuki models
\cite{Borisov:1997nc,Bantay:1999us}.

To get an idea how much is still missing, note that
we encountered ``only" \totBettis~different Hodge number pairs, while
for example in \cite{Kreuzer:2000xy}
already more than 30,000 are presented.

For any accessible RCFT one may consider other brane constructions than the four stack
model, for example a construction that would yield $SU(5)$ unification at the string
scale, or broken versions of such a model. This would have the obvious advantage of
restoring gauge unification.
One could also look for Pati-Salam like models,
and for models that require a brane recombination mechanism to yield the standard model.
As mentioned in the introduction, examples of these types have been constructed already.
In principle there is no reason why the standard model could not be realized in such
a way, but of course this opens a Pandora's box of possibilities. The number of solutions
would explode even more if we were to allow chiral exotics, a possibility that can never be
rigorously ruled out experimentally.
But if indeed the quarks and leptons we know today provide misleading information about the
true chiral spectrum of nature, then the goal of finding the standard model in string theory
is essentially unachievable at present.

The obvious way to control this explosion of candidate standard models
is to apply more constraints. There are several criteria
that might be used to reject some of the many models on our list, such as absence of a Higgs
candidate, the absence of plausible mechanisms to break $Sp(2)_{\rm c}$, or to generate a
$B-L$ gauge boson mass. However, it is hard to come up with a criterium that is at the same
time very effective and unquestionable.

One could also try to work out the Yukawa coupling of the standard model particles
to the Higgs sector. This requires open string three-point couplings, computable in
principle in RCFT, but not yet in practice. In cases with more than one supersymmetric Higgs
pair, there is the obvious problem of deciding which combination gets a vacuum expectation value.
This could be treated as a set of free parameters, and fitted to the observed couplings. Doing that would
requires a renormalization group evolution of those couplings, which in its turn requires
detailed knowledge of the full (chiral and non-chiral) light spectrum. Then one still faces the
problem of the moduli-dependence of the result. To some extent, this may be studied by probing
some directions in moduli space with a large number of rational points, as we did for the gauge
couplings. However, such an effort is probably premature. Most likely an essential ingredient
is missing for the understanding of Yukawa couplings, and that is especially true for neutrino masses.

Fortunately more experimental constraints can be expected in the coming years, especially from the
LHC, and also from neutrino and astrophysics experiments. With more experimental input we might come to the
conclusion that another type of model should be searched for, and with a different
set of constraints. It is to be expected that the class
considered here will yield equally rich and abundant results in other cases.\\
\vspace{1cm}

\noindent{\bf Acknowledgements:}\\
We would like to thank Dieter L\"ust for suggesting a check
of the results of \cite{Blumenhagen:2003jy}. We would like to
thank Jeff Templon, David Groep, and Davide Salomoni for assistence with
running jobs on the NIKHEF computer farm.
The work of A.N.S. has been performed as part of the program
FP 52 of the Foundation for Fundamental Research of Matter (FOM), and
the work of T.P.T.D. and A.N.S. has been performed as part of the
program FP 57 of FOM. The work A.N.S. has been partially
supported by funding of the Spanish ``Ministerio de
Ciencia y Tecnolog\'\i a", Project BFM2002-03610.
The work of L.R.H. wass supported in part by the Federal Office for Scientific,
Technical and Cultural Affairs through the
"Interuniversity Attraction Poles Programme -- Belgian Science Policy"
P5/27.


\section{Appendix: Boundary and Ishibashi orbits}\label{borbit} 

To compute all annuli using the formalism of \cite{FHSSW} it is
sufficient to compute all reflection coefficients, and use (\ref{Acoef}), summing
over all Ishibashi labels to compute all annuli $A^i_{~ab}$. Computationally
this scales as $N_i (N_B)^3$, where $N_i$ is the number of primaries one needs
to consider (in our case only the massless ones) and $N_B$ the number of boundaries,
or equivalently, the number of Ishibashi labels. This sort of computation is
analogous to the computation of the fusion coefficients using the Verlinde formula,
which, for $N$ primaries, scales as $N^4$. In the latter case, such computations can
be speeded up drastically if there is a non-trivial simple current group ${\cal G}$. 
This leads to relations among the matrix elements of the matrix $S$ of the form
\beq \label{eq:SCS}
 S_{Ji,j}=e^{2\pi i Q_J(j)}S_{ij} \ .
\eeq
Fusion computations can then be made more efficient
 for the
following three reasons
\begin{enumerate}
\item Simple current charge conservation, {\it i.e.} \\ \mbox{$N_{abc}=0$ if $Q_J(a)\!+Q_J(b)\!+Q_J(c) \not= 0\mod 1$}.
\item Simple current relations among fusion entries, $N_{Ka,Jb,Lc}=N_{abc}$ if $JL=K$
\item The summation in the Verlinde formula can be restricted to orbit representatives.
\end{enumerate}
In the absence of fixed points, {\it i.e} if all orbit are equally long, this reduces the
size of the computation by a factor $|{\cal G}|^4$ to $(N_O)^4$, where $N_O$ is the number of orbits.

In the present case the quantity of interest is $A^i_{~ab}$, for a subset of labels $i$.  In order
to make use of similar results, we have to establish an action of simple currents on the boundary labels
$a$ as well as an action on the Ishibashi labels in the internal summation. 

First we define a boundary simple current. Consider the oriented annulus
$$ A_{ia}^{~~b}  = \sum_m  {S_{im}\over S_{0m}} R_{ma} R_{mb}^* \ , $$
where for simplicity the boundary and Ishibashi labels $a$ and $m$ implicitly include the degeneracy label.
The labels $i$ are primaries of the bulk CFT, and we can consider the case
that $i$ is a simple current, $J$. Then
one can show that for every $a$ there is precisely one $b$ such that $A_{Ja}^{~~~}b=1$, and that
for all other $b$ this quantity vanishes.
The proof goes as follows.
$$ A_{Ja}^{~~~b}= \sum_m  {S_{Jm}\over S_{0m}} R_{ma} R_{mb}^* =  \sum_m  e^{2\pi i Q_J(m)} R_{ma} R_{mb}^* $$
For each $a$, the vectors $R_{ma}$ have norm 1. Hence this sum can at most be 1, and it is equal to 1 if
and only if
\beq \label{eq:SCP}
 R_{mb}=e^{2\pi i Q_J(m)} R_{ma}
\eeq
Since the boundaries are independent, there can be at most one boundary $b$ that satisfies this.
and
since the set is complete, there must be precisely one (note that the matrix $A_J$ has to satisfy the
completeness condition $A_J A_{J^c} = {\bf 1}$ and hence cannot have an entire row that is zero).
Hence it makes sense to define $J\cdot a \equiv b$ as
the action of $J$ on $a$. Using (\ref{eq:SCP}) one can easily show that
$$ A_{i,J\cdot a}^{~~J \cdot b} = A_{ia}^{~~b}\ .$$
This establishes the second of the three properties listed above, with $K=1$. Since the label $i$ is restricted
to massless fields a further generalization analogous to $K\not=1$ is not really useful, since the simple
current action on $i$ does not preserve its conformal weight.

A simple current $J$ may yield a trivial action on all boundaries. It is easy to show
that $J\cdot(K\cdot a)=(JK \cdot a)= K\cdot(J\cdot a)$, where $JK$ denotes the fusion product of the CFT. Therefore
if $J$ and $K$ both fix all boundaries, so does $JK$, and hence the set of simple currents that fix all boundaries
is a subgroup ${\cal G}_{X}$ of ${\cal G }$.  The boundary simple current group is defined as the
discrete group ${\cal G } / {\cal G}_{X}$. The elements of that group have distinct actions on at least one
of the boundaries.

In all cases covered by the simple current formalism of \cite{FHSSW} there is a special boundary
originating from the CFT vacuum state, $a=0$.  One might conjecture a natural correspondence
between the elements of the boundary simple current group and the set of boundaries
obtained by the action of ${\cal G}$ on the boundary 0. However, it turns
out that this is not a one-to-one correspondence: it may happen that $J\cdot 0 = 0$, but
$J\cdot a \not=a$ for some other $a$. Hence the boundary simple current group may be larger
than the set of ``simple current boundaries".

Relation (\ref{eq:SCP}) is the analog of (\ref{eq:SCS}) for the action on the boundary label.
An action on the Ishibashi label can be derived directly from (\ref{eq:BRC}) and  (\ref{eq:FPact}):
\beq \label{eq:IshiAct}
R_{(Im,J)[a,\psi_a]}=F_m(I,J) e^{2\pi i Q_I(a)}R_{(m,J)[a,\psi_a]}  \ ,
\eeq
if the set of degeneracy labels $J$ for $m$ and $Lm$ are the same. This is the
case if and only if $L$ is local with respect to the simple current group $\cH$ that
defines the MIPF. The subgroup of $\cG$ with that property will be called the Ishibashi simple
current group. The simple current twist $F_m(I,J)$ is a sign in all known cases, and this
implies that its occurrence here is irrelevant, since we are free to change all boundary and crosscap
coefficients simultaneously by signs: $R_{ma} \to \epsilon_m R_{ma}, U_m \to \epsilon_m U_{m} $. This
allows us to redefine the boundary coefficients on each Ishibashi orbit in such a way that  for the
orbit representatives, $m_0$, we get simply
\beq 
R_{(Im_0,J)[a,\psi_a]}= e^{2\pi i Q_I(a)}R_{(m_0,J)[a,\psi_a]}  \ ,
\eeq
Note that the crosscap coefficients $U_{m,J}$ vanish if $J\not=0$, so these sign changes are
irrelevant for crosscaps.

Now consider the first simplification listed above, charge conservation.
The annulus amplitude $A^i_{~ab}$ can be expressed in terms of the
fixed point fusion coefficients \cite{thesis_lennaert}
$$ (N^J)^j_{ab} = \sum_m {S_{jm}^* \over S_{0m}} S^J_{ma}S^{J^c}_{mb} \ . $$
 as a linear combination of the form
$$ \sum_{L \in \cH}\sum_{J \in cH} \phi(L,J,a,b) (N^J)^{KLi}_{ab}\ , $$
Here $\phi(L,J,a,b)$ are complex coefficients and $K$ is the Klein bottle current.
Using \ref{eq:FPact} and the property $F_a(K,N)F_a(K,M)=F_a(K,NM)$ it follows that
 $ (N^J)^j_{ab} = 0 $ unless $Q_M(a)+Q_M(b)=Q_M(j) \mod 1$ for all simple currents $M$.
For the annulus this implies that $A^i_{~ab} = 0$ unless there is an $L \in \cH$
so that $Q_M(a)+Q_M(b)\not= Q_M(i)+Q_M(K)+Q_M(L)\mod 1$.

Finally consider the third simplification. The sum over $m$ can be reduced to a sum over
Ishibashi orbit representatives provided that charge conservation is satisfied,
{\it i.e} $Q_I(a)+Q_I(b)\not= Q_I(i)+Q_I(K)+Q_I(L)\mod 1$, for all $L \in \cH$. Since
Ishibashi currents $I$ are always local with respect to $\cH$, this condition is always
satisfied.

\section{Bibliography}
\bibliography{REFS}
\bibliographystyle{h-elsevier}

\newpage
\section{Tables}
\footnotesize

\begin{table}[h]\caption{List of standard model representations that can appear, and their labelling.}\label{tbl:sm_reps}

\begin{center}
\footnotesize
~~~~~~~~~~~~~~~~\begin{tabular}{|l|c|cl|cl|c|c|} \hline
nr. &  $U(3)_{\rm a}$ & $U(2)_{\rm b}$ & $Sp(2)_{\rm b}$ & $U(1)_{\rm c} $ &  $SO(2)_{\rm c}$ & $U(1)_{\rm d}$  & massless particle  \\
& & & & & $Sp(2)_{\rm c}$ & &\\
\hline \hline
1  & $V$    & $V $  & $\left.\right\rmoustache$ \raisebox{-1.5\height}[0cm][0cm]{$V$}  &  0&	 & 0	 &  $(u,d)$  \\
2  & $V$    & $V^*$ & $\left.\right\lmoustache$ &  0&	 & 0	 &  $(u,d)$  \\
3  & $V^*$  & 0	    & &  $V$	 &$\left.\right\rmoustache$ \raisebox{-1.5\height}[0cm][0cm]{$V$} & 0	  &  $u^c$  \\
4  & $V^*$  & 0	    & &  $V^*$	 &$\left.\right\lmoustache$ & 0	  &  $d^c$  \\
5  & 0	    & $V$   & $\left.\right\rmoustache$ \raisebox{-1.5\height}[0cm][0cm]{$V$} &  0&	 & $V$	 &  $(\nu,e^{-})$  \\
6  & 0	    & $V^*$ & $\left.\right\lmoustache$&  0&	 & $V$	 &  $(\nu,e^{-})$  \\
7  & 0	    & 0	    & &  $V$	 &$\left.\right\rmoustache$ \raisebox{-1.5\height}[0cm][0cm]{$V$} & $V^*$  & $\nu^c$  \\
8  & 0	    & 0	    & &  $V^*$	 &$\left.\right\lmoustache$ & $V^*$  & $e^{+}$  \\  \hline
9  & 0	    & $V$   & &  $V$&	 & 0	  & $H_1$  \\
10 & 0	    & $V$   & &  $V^*$&	 & 0	  & $H_2$  \\ \hline
11 & $V$    & 0	    & &  0&	 & $V$	  & $(3,1,-\frac13)_{1,1}$  \\
12 & $V$    & 0	    & &  0&	 & $V^*$  & $(3,1,\frac23)_{1,-1}$  \\
13 & Adj    & 0	    & &  0&	 & 0	  & $(8,1,0)_{0,0}+(1,1,0)_{0,0}$  \\
14 & A	    & 0	    & &  0&	 & 0	  & $(3^*,1,\frac13)_{2,0}$  \\
15 & S	    & 0	    & &  0&	 & 0	  & $(6,1,\frac13)_{2,0}$  \\
16 & 0	    & Adj   & --- &  0&	 & 0	  & $(1,3,0)_{0,0}+(1,1,0)_{0,0}$  \\
17 & 0	    & A	    & &  0&	 & 0	  & $(1,1,0)_{0,0}$  \\
18 & 0	    & S	    & &  0&	 & 0	  & $(1,3,0)_{0,0}$  \\
19 & 0	    & 0	    & &  Adj & ---	 & 0	  & $(1,1,0)_{0,0}$  \\
20 & 0	    & 0	    & &  --- &A	 & 0	  & $(1,1,-1)_{0,0}$  \\
21 & 0	    & 0	    & &  S&	 & 0	  & $(1,1,-1)_{0,0}$  \\
22 & 0	    & 0	    & &  0&	 & Adj	  & $(1,1,0)_{0,0}$  \\
23 & 0	    & 0	    & &  0&	 & A	  & ---  \\
24 & 0	    & 0	    & &  0&	 & S	  & $(1,1,-1)_{0,2}$  \\ \hline
25 & $V$    & 0	    & &  0&	 & 0	  & $(3,1,\frac16)_{1,0}$  \\
26 & 0	    & $V$   & &  0&	 & 0	  & $(1,2,0)_{0,0}$  \\
27 & 0	    & 0	    & &  $V$&	 & 0	  & $(1,1,-\frac12)_{0,0}$  \\
28 & 0	    & 0	    & &  0&	 & $V$	  &  $(1,1,-\frac12)_{0,1}$  \\ \hline
29 & 0	    & 0	    & &  0&	 & $0$	  &  $(1,1,0)_{0,0}$ \\ \hline
\end{tabular}

\end{center}
\end{table}

\begin{longtable}{|l|c|c|c|c|c|c|c|}
\caption{Summary results for all Gepner Models}\\
\hline 
nr. &  Tensor & Prim & S.C. & MIPF  & Intersect. & Sol. & SM spectra \\
\hline
\endhead
\hline 
\endfoot 
\input{gepner_table.inc}
\label{tbl:gepner_models}
\end{longtable}

\begin{longtable}{|l|c|c|c|c|}
\caption{All MIPFs with solutions}\\
\hline
Tensor &  $(h_{11},h_{21},S) $ & Boundaries  & Nr. & Types \\
\hline
\endhead
\hline 
\endfoot 
\input{mipfs_with_sols.inc} 
\label{tbl:mipfs_sols}
\end{longtable}

\begin{table}\caption{Distribution of chiral standard model types, distinguished
by CP group and the $U(2)_{\rm b}$ anomaly. In the ``Higgs" column, "$-2n$"
implies that there a $n$ supersymmetric Higgs pairs $(2,\frac12)+(2,-\frac12)$
that are chiral with respect to $U(2)_{\rm b}$ (the sign is just a convention).}\label{tbl:nr_sols}

\begin{center}
\begin{tabular}{|l|c|c|c|c|}\hline
Type & Quarks & Leptons & Higgs & Total \\ \hline 
\input{chiral_types_table.inc}
\hline
\end{tabular}
\end{center}
\end{table}

\normalsize

\end{document}

%% file: distribution.inc
\newcommand{\percMasslessBminL}{3}

\newcommand{\noMirrors}{2018}

\newcommand{\maxHiggs}{56}

\newcommand{\masslessBminLandNoBranes}{22}

%% file: hiddenbranes.inc
\newcommand{\maxdim}{780}
\newcommand{\maxSO}{18}
\newcommand{\maxU}{18}
\newcommand{\maxSP}{24}

%% file: famstatistics.inc
\newcommand{\mipfsFaMonetonine}{18}
\newcommand{\mipfsFaMthreetonine}{78}

%% file: statistics.inc
\newcommand{\totMipfs}{5403}

\newcommand{\totBettis}{873}
\newcommand{\totHeterotic}{1829}

\newcommand{\mipfsWOinter}{4079}
\newcommand{\mipfsWsols}{333}
\newcommand{\mipfsWaborted}{495}
\newcommand{\mipfsWsolsWaborted}{153}
\newcommand{\totInters}{45051902}
\newcommand{\intersWsols}{1635985}

\newcommand{\totSpectra}{179520}
\newcommand{\mipfsWmissTypes}{20}
\newcommand{\absentMipfs}{2}

%% file: gepner_table.inc
1 & (1,5,41,1804) & 28539 & 1 & 1,0,0 & 0 & 0 & 0 \\
2 & (1,5,42,922) & 29772 & 2 & 2,-,- & - & - & - \\
3 & (1,5,43,628) & 31482 & 3 & 2,-,- & - & - & - \\
4 & (1,5,44,481) & 9399 & 1 & 1,0,0 & 0 & 0 & 0 \\
5 & (1,5,46,334) $\dagger$& 35442 & 6 & 4,1,0 & 12 & 0 & 0 \\
6 & (1,5,47,292) $\dagger$& 37800 & 7 & 2(1),1,0 & 1128 & 0 & 0 \\
7 & (1,5,49,236) & 4575 & 1 & 1,0,0 & 0 & 0 & 0 \\
8 & (1,5,52,187) & 12690 & 3 & 2(1),1,0 & 144 & 0 & 0 \\
9 & (1,5,54,166) $\dagger$& 48258 & 14 & 4,1,0 & 54 & 0 & 0 \\
10 & (1,5,58,138) & 6156 & 2 & 2,0,0 & 0 & 0 & 0 \\
11 & (1,5,61,124) & 64449 & 21 & 4(2),3,0 & 81044 & 0 & 0 \\
12 & (1,5,68,103) & 20748 & 7 & 2(1),1,0 & 234 & 0 & 0 \\
13 & (1,5,76,89) & 2835 & 1 & 1,0,0 & 0 & 0 & 0 \\
14 & (1,5,82,82) $\dagger$& 108612 & 42 & 8(1),3,0 & 1744 & 0 & 0 \\
15 & (1,6,23,598) & 12600 & 2 & 2,0,0 & 0 & 0 & 0 \\
16 & (1,6,24,310) $\dagger$& 13650 & 4 & 6(1),1,0 & 18260 & 0 & 0 \\
17 & (1,6,25,214) & 14742 & 6 & 4(2),2,0 & 212000 & 0 & 0 \\
18 & (1,6,26,166) & 15876 & 8 & 10(7),8,0 & 4939262 & 0 & 0 \\
19 & (1,6,28,118) & 18270 & 12 & 12(9),10,2 & 1289765 & 217 & 67 \\
20 & (1,6,30,94) & 20664 & 16 & 14(12),12,1 & 2912087 & 297 & 221 \\
21 & (1,6,31,86) & 2464 & 2 & 2,0,0 & 0 & 0 & 0 \\
22 & (1,6,34,70) & 26460 & 24 & 20(16),18,4 & 1539069 & 2466 & 1587 \\
23 & (1,6,38,58) & 8260 & 8 & 10(7),8,1 & 435815 & 36 & 18 \\
24 & (1,6,40,54) & 4018 & 4 & 6(3),3,0 & 87494 & 0 & 0 \\
25 & (1,6,46,46) & 46536 & 48 & 28(24),26,5 & 3687098 & 67723 & 16490 \\
26 & (1,7,17,340) & 9396 & 3 & 2,0,0 & 0 & 0 & 0 \\
27 & (1,7,18,178) & 10224 & 6 & 4(1),1,0 & 29672 & 0 & 0 \\
28 & (1,7,19,124) & 11880 & 9 & 6(2),4,0 & 2513 & 0 & 0 \\
29 & (1,7,20,97) & 4116 & 3 & 2,0,0 & 0 & 0 & 0 \\
30 & (1,7,22,70) & 14760 & 18 & 12(6),8,0 & 38210 & 0 & 0 \\
31 & (1,7,25,52) & 21060 & 27 & 10(2),7,0 & 7523 & 0 & 0 \\
32 & (1,7,28,43) & 7128 & 9 & 6,2,0 & 19 & 0 & 0 \\
33 & (1,7,34,34) & 33264 & 54 & 16(8),10,2 & 62992 & 70 & 13 \\
34 & (1,8,14,238) & 8082 & 4 & 6(3),3,1 & 3447270 & 16 & 16 \\
35 & (1,8,16,88) & 10218 & 12 & 12(7),7,0 & 70962 & 0 & 0 \\
36 & (1,8,18,58) & 12690 & 20 & 12(7),9,1 & 240804 & 4 & 2 \\
37 & (1,8,22,38) & 2034 & 4 & 6(1),1,0 & 2888 & 0 & 0 \\
38 & (1,8,28,28) & 28410 & 60 & 20(8),14,2 & 152706 & 1288 & 188 \\
39 & (1,9,12,229) & 2875 & 1 & 1,0,0 & 0 & 0 & 0 \\
40 & (1,9,13,108) & 1015 & 1 & 1,0,0 & 0 & 0 & 0 \\
41 & (1,9,20,31) & 6160 & 11 & 2,0,0 & 0 & 0 & 0 \\
42 & (1,10,11,154) & 7704 & 6 & 4,1,0 & 372 & 0 & 0 \\
43 & (1,10,12,82) & 8970 & 12 & 12(4),9,0 & 277382 & 0 & 0 \\
44 & (1,10,13,58) & 10332 & 18 & 12(4),9,0 & 19649 & 0 & 0 \\
45 & (1,10,14,46) & 11748 & 24 & 20(12),17,3 & 328229 & 7389 & 3687 \\
46 & (1,10,16,34) & 14994 & 36 & 36(18),29,5 & 198765 & 503 & 173 \\
47 & (1,10,18,28) & 4698 & 12 & 12(3),6,2 & 19344 & 32 & 14 \\
48 & (1,10,19,26) & 2280 & 6 & 4,0,0 & 0 & 0 & 0 \\
49 & (1,10,22,22) & 26532 & 72 & 50(20),39,10 & 499730 & 15055 & 2780 \\
50 & (1,11,11,76) & 9828 & 13 & 2,1,0 & 96 & 0 & 0 \\
51 & (1,12,12,40) & 12138 & 28 & 10(2),8,1 & 22942 & 8 & 2 \\
52 & (1,12,13,33) & 595 & 1 & 1,0,0 & 0 & 0 & 0 \\
53 & (1,12,19,19) & 10500 & 21 & 4,0,0 & 0 & 0 & 0 \\
54 & (1,13,13,28) & 19845 & 45 & 10,5,0 & 1854 & 0 & 0 \\
55 & (1,13,18,18) & 3220 & 10 & 4,1,0 & 20 & 0 & 0 \\
56 & (1,14,14,22) & 5336 & 16 & 14(1),10,2 & 17112 & 63 & 23 \\
57 & (1,16,16,16) & 33210 & 108 & 26(5),17,5 & 125476 & 9204 & 494 \\
58 & (2,3,19,418) & 6320 & 2 & 2,0,0 & 0 & 0 & 0 \\
59 & (2,3,20,218) & 6972 & 4 & 6(4),4,0 & 315210 & 0 & 0 \\
60 & (2,3,22,118) & 8232 & 8 & 5(4),4,1 & 180431 & 6 & 4 \\
61 & (2,3,23,98) & 9120 & 10 & 4(1),1,0 & 1204 & 0 & 0 \\
62 & (2,3,26,68) & 3036 & 4 & 6(4),4,0 & 18416 & 0 & 0 \\
63 & (2,3,28,58) & 13340 & 20 & 12(8),9,4 & 142742 & 482 & 176 \\
64 & (2,3,34,43) & 1232 & 2 & 2,0,0 & 0 & 0 & 0 \\
65 & (2,3,38,38) & 22920 & 40 & 10(8),9,2 & 151792 & 1044 & 216 \\
66 & (2,4,11,154) & 3540 & 4 & 6(4),4,0 & 2304160 & 0 & 0 \\
67 & (2,4,12,82) & 4160 & 8 & 30(26),26,12 & 5133558 & 598 & 294 \\
68 & (2,4,13,58) & 4830 & 12 & 12(7),9,1 & 468648 & 146 & 69 \\
69 & (2,4,14,46) & 5340 & 16 & 27(24),25,13 & 1918601 & 17411 & 5607 \\
70 & (2,4,16,34) & 7140 & 24 & 60(48),54,23 & 3700006 & 218598 & 45055 \\
71 & (2,4,18,28) & 2320 & 8 & 30(23),25,15 & 745644 & 801 & 360 \\
72 & (2,4,19,26) & 1100 & 4 & 6(2),3,0 & 23872 & 0 & 0 \\
73 & (2,4,22,22) & 12060 & 48 & 54(39),51,25 & 3403934 & 423560 & 43532 \\
74 & (2,5,8,138) & 2862 & 4 & 6(3),4,0 & 191424 & 0 & 0 \\
75 & (2,5,10,40) & 1230 & 4 & 6(1),3,0 & 5502 & 0 & 0 \\
76 & (2,5,12,26) & 6006 & 28 & 12,7,3 & 34744 & 2426 & 431 \\
77 & (2,6,7,70) & 3024 & 8 & 5(1),3,0 & 28234 & 0 & 0 \\
78 & (2,6,8,38) & 3780 & 16 & 27(18),22,7 & 323662 & 6313 & 1368 \\
79 & (2,6,10,22) & 5544 & 32 & 59(20),53,27 & 361546 & 18964 & 3624 \\
80 & (2,6,14,14) & 9744 & 64 & 87(20),71,30 & 758636 & 62856 & 5424 \\
81 & (2,7,7,34) & 4032 & 18 & 6,0,0 & 0 & 0 & 0 \\
82 & (2,7,10,16) & 2040 & 12 & 12,6,1 & 10504 & 4 & 1 \\
83 & (2,8,8,18) & 6480 & 40 & 44(3),32,16 & 1019592 & 222006 & 17311 \\
84 & (2,8,10,13) & 630 & 4 & 6,3,0 & 1320 & 0 & 0 \\
85 & (2,10,10,10) & 12000 & 96 & 92(7),71,34 & 850844 & 137472 & 9878 \\
86 & (3,3,9,108) & 2900 & 5 & 2,0,0 & 0 & 0 & 0 \\
87 & (3,3,10,58) & 3280 & 10 & 4,1,0 & 124 & 0 & 0 \\
88 & (3,3,12,33) & 1700 & 5 & 2,0,0 & 0 & 0 & 0 \\
89 & (3,3,13,28) & 6300 & 25 & 6,2,0 & 118 & 0 & 0 \\
90 & (3,3,18,18) & 9200 & 50 & 12,4,0 & 1492 & 0 & 0 \\
91 & (3,4,6,118) & 2100 & 4 & 6(3),3,0 & 176520 & 0 & 0 \\
92 & (3,4,7,43) & 1584 & 3 & 2,0,0 & 0 & 0 & 0 \\
93 & (3,4,8,28) & 3280 & 20 & 12(1),7,0 & 3348 & 0 & 0 \\
94 & (3,4,10,18) & 540 & 4 & 6,0,0 & 0 & 0 & 0 \\
95 & (3,4,13,13) & 4410 & 15 & 4,0,0 & 0 & 0 & 0 \\
96 & (3,5,5,68) & 2394 & 7 & 2,0,0 & 0 & 0 & 0 \\
97 & (3,6,6,18) & 1064 & 8 & 10,1,0 & 88 & 0 & 0 \\
98 & (3,8,8,8) & 9200 & 100 & 18,11,2 & 5406 & 96 & 4 \\
99 & (4,4,5,40) & 2322 & 12 & 10,6,0 & 56000 & 0 & 0 \\
100 & (4,4,6,22) & 3150 & 24 & 44(9),33,15 & 465222 & 51448 & 8737 \\
101 & (4,4,7,16) & 3888 & 36 & 26(1),17,6 & 93764 & 2590 & 59 \\
102 & (4,4,8,13) & 1218 & 12 & 10(2),7,0 & 3682 & 0 & 0 \\
103 & (4,4,10,10) & 7200 & 72 & 110(3),74,15 & 999730 & 277752 & 9983 \\
104 & (4,5,5,19) & 1890 & 7 & 2,0,0 & 0 & 0 & 0 \\
105 & (4,6,6,10) & 1540 & 16 & 54,22,3 & 6874 & 64 & 6 \\
106 & (4,7,7,7) & 5184 & 27 & 6,0,0 & 0 & 0 & 0 \\
107 & (5,5,5,12) & 6615 & 49 & 5,0,0 & 0 & 0 & 0 \\
108 & (6,6,6,6) & 9632 & 128 & 76,44,10 & 174232 & 70864 & 1310 \\
109 & (1,1,2,11,154) & 2124 & 6 & 4,0,0 & 0 & 0 & 0 \\
110 & (1,1,2,12,82) & 2496 & 12 & 12,0,0 & 0 & 0 & 0 \\
111 & (1,1,2,13,58) & 2898 & 18 & 10,0,0 & 0 & 0 & 0 \\
112 & (1,1,2,14,46) & 3204 & 24 & 10,2,0 & 788 & 0 & 0 \\
113 & (1,1,2,16,34) & 4284 & 36 & 30,8,0 & 260 & 0 & 0 \\
114 & (1,1,2,18,28) & 1392 & 12 & 12,0,0 & 0 & 0 & 0 \\
115 & (1,1,2,19,26) & 660 & 6 & 4,0,0 & 0 & 0 & 0 \\
116 & (1,1,2,22,22) & 7236 & 72 & 26,7,0 & 3864 & 0 & 0 \\
117 & (1,1,3,6,118) & 1260 & 6 & 4,0,0 & 0 & 0 & 0 \\
118 & (1,1,3,7,43) & 1584 & 9 & 5,0,0 & 0 & 0 & 0 \\
119 & (1,1,3,8,28) & 2010 & 30 & 8,1,0 & 2 & 0 & 0 \\
120 & (1,1,3,10,18) & 324 & 6 & 4,0,0 & 0 & 0 & 0 \\
121 & (1,1,3,13,13) & 4410 & 45 & 10,1,0 & 54 & 0 & 0 \\
122 & (1,1,4,5,40) & 1431 & 18 & 10,0,0 & 0 & 0 & 0 \\
123 & (1,1,4,6,22) & 1890 & 36 & 30,0,0 & 0 & 0 & 0 \\
124 & (1,1,4,7,16) & 2484 & 54 & 40,1,0 & 2 & 0 & 0 \\
125 & (1,1,4,8,13) & 819 & 18 & 10,0,0 & 0 & 0 & 0 \\
126 & (1,1,4,10,10) & 4320 & 108 & 96,6,0 & 652 & 0 & 0 \\
127 & (1,1,5,5,19) & 1890 & 21 & 4,0,0 & 0 & 0 & 0 \\
128 & (1,1,6,6,10) & 924 & 24 & 20,1,0 & 16 & 0 & 0 \\
129 & (1,1,7,7,7) & 5184 & 81 & 19,2,0 & 8 & 0 & 0 \\
130 & (1,2,2,5,40) & 492 & 4 & 6,0,0 & 0 & 0 & 0 \\
131 & (1,2,2,6,22) & 1512 & 32 & 31,7,0 & 3434 & 0 & 0 \\
132 & (1,2,2,7,16) & 816 & 12 & 12,0,0 & 0 & 0 & 0 \\
133 & (1,2,2,8,13) & 252 & 4 & 6,0,0 & 0 & 0 & 0 \\
134 & (1,2,2,10,10) & 3288 & 96 & 120,22,1 & 7566 & 72 & 2 \\
135 & (1,2,3,3,58) & 920 & 10 & 4,0,0 & 0 & 0 & 0 \\
136 & (1,2,3,4,18) & 160 & 4 & 6,0,0 & 0 & 0 & 0 \\
137 & (1,2,4,4,10) & 2250 & 72 & 118,19,4 & 10920 & 730 & 99 \\
138 & (1,2,4,6,6) & 420 & 16 & 27,0,0 & 0 & 0 & 0 \\
139 & (1,3,3,3,13) & 1400 & 25 & 4,0,0 & 0 & 0 & 0 \\
140 & (1,3,3,4,8) & 260 & 10 & 4,0,0 & 0 & 0 & 0 \\
141 & (1,4,4,4,4) & 4266 & 216 & 112,11,2 & 5552 & 172 & 6 \\
142 & (2,2,2,3,18) & 1040 & 32 & 27,2,0 & 1168 & 0 & 0 \\
143 & (2,2,2,4,10) & 1520 & 64 & 230,60,7 & 48876 & 4832 & 92 \\
144 & (2,2,2,6,6) & 3024 & 128 & 305,101,6 & 111080 & 10304 & 95 \\
145 & (2,2,3,3,8) & 720 & 20 & 12,0,0 & 0 & 0 & 0 \\
146 & (2,2,4,4,4) & 1500 & 48 & 180,5,0 & 4640 & 0 & 0 \\
147 & (3,3,3,3,3) & 4000 & 125 & 8,0,0 & 0 & 0 & 0 \\
148 & (1,1,1,1,5,40) & 972 & 27 & 8,0,0 & 0 & 0 & 0 \\
149 & (1,1,1,1,6,22) & 1134 & 54 & 16,0,0 & 0 & 0 & 0 \\
150 & (1,1,1,1,7,16) & 1944 & 81 & 34,1,1 & 6 & 3 & 2 \\
151 & (1,1,1,1,8,13) & 756 & 27 & 8,0,0 & 0 & 0 & 0 \\
152 & (1,1,1,1,10,10) & 2592 & 162 & 58,0,0 & 0 & 0 & 0 \\
153 & (1,1,1,2,3,18) & 96 & 6 & 4,0,0 & 0 & 0 & 0 \\
154 & (1,1,1,2,4,10) & 1350 & 108 & 72,0,0 & 0 & 0 & 0 \\
155 & (1,1,1,2,6,6) & 252 & 24 & 10,0,0 & 0 & 0 & 0 \\
156 & (1,1,1,3,3,8) & 240 & 15 & 4,0,0 & 0 & 0 & 0 \\
157 & (1,1,1,4,4,4) & 2673 & 324 & 142,0,0 & 0 & 0 & 0 \\
158 & (1,1,2,2,2,10) & 912 & 96 & 52,0,0 & 0 & 0 & 0 \\
159 & (1,1,2,2,4,4) & 900 & 72 & 110,0,0 & 0 & 0 & 0 \\
160 & (1,2,2,2,2,4) & 440 & 64 & 138,0,0 & 0 & 0 & 0 \\
161 & (2,2,2,2,2,2) & 2944 & 512 & 1031,10,0 & 448 & 0 & 0 \\
162 & (1,1,1,1,1,2,10) & 810 & 162 & 34,0,0 & 0 & 0 & 0 \\
163 & (1,1,1,1,1,4,4) & 1944 & 486 & 156,0,0 & 0 & 0 & 0 \\
164 & (1,1,1,1,2,2,4) & 540 & 108 & 48,0,0 & 0 & 0 & 0 \\
165 & (1,1,1,2,2,2,2) & 264 & 96 & 46,0,0 & 0 & 0 & 0 \\
166 & (1,1,1,1,1,1,1,4) & 2187 & 729 & 124,0,0 & 0 & 0 & 0 \\
167 & (1,1,1,1,1,1,2,2) & 324 & 162 & 24,0,0 & 0 & 0 & 0 \\
168 & (1,1,1,1,1,1,1,1,1) & 2187 & 2187 & 152,0,0 & 0 & 0 & 0 

%% file: mipfs_with_sols.inc
        (1,6,28,118) & (24,84,429) & 2400 & 6 &  (6,0,0,0,50,0,0) \\
                       & (75,75,565) & 6090 & 2 &  (2,0,0,2,15,0,0) \\
         (1,6,30,94) & (24,84,425) & 1980 & 6 &  (6,0,0,0,221,0,0) \\
         (1,6,34,70) & (29,125,557) & 3312 & 10 &  (10,0,3,2,0,0,0) \\
                       & (14,98,451) & 1656 & 9 &  (9,0,0,0,1557,0,0) \\
                       & (43,91,509) & 4464 & 2 &  (2,0,0,2,11,0,0) \\
                       & (29,53,353) & 3420 & 11 &  (11,0,0,0,6,0,0) \\
         (1,6,38,58) & (29,53,351) & 1520 & 6 &  (6,0,0,0,18,0,0) \\
         (1,6,46,46) & (19,163,649) & 2968 & 12 &  (12,0,87,0,11,0,0) \\
                       & (9,129,525) & 1484 & 10 &  (10,0,0,0,16294,0,0) \\
                       & (27,123,569) & 3944 & 2 &  (2,0,4,20,37,0,0) \\
                       & (19,67,377) & 2968 & 13 &  (13,0,6,1,29,0,0) \\
                       & (59,59,449) & 7888 & 4 &  (4,0,0,0,1,0,0) \\
         (1,7,34,34) & (23,95,437) & 2136 & 10 &  (10,0,0,2,0,0,0) \\
                       & (29,77,457) & 3696 & 2 &  (2,0,7,0,4,0,0) \\
        (1,8,14,238) & (53,53,443) & 3252 & 4 &  (4,0,0,0,16,0,0) \\
         (1,8,18,58) & (38,62,377) & 2538 & 2 &  (2,0,0,0,2,0,0) \\
         (1,8,28,28) & (17,95,419) & 1894 & 2 &  (2,0,110,20,54,0,0) \\
                       & (29,29,251) & 2316 & 12 &  (12,0,4,0,0,0,0) \\
        (1,10,14,46) & (8,68,315) & 728 & 8 &  (8,0,0,0,3447,0,0) \\
                       & (17,41,259) & 1540 & 11 &  (11,0,123,0,92,0,0) \\
                       & (35,59,369) & 2000 & 2 &  (2,0,0,16,6,0,0) \\
        (1,10,16,34) & (13,97,405) & 1088 & 20 &  (20,0,8,0,16,0,0) \\
                       & (29,65,355) & 1666 & 22 &  (22,0,20,42,44,10,0) \\
                       & (20,32,241) & 1920 & 25 &  (25,0,6,0,0,0,0) \\
                       & (20,32,249) & 1920 & 29 &  (29,0,9,0,12,0,0) \\
                       & (38,38,311) & 2544 & 7 &  (7,0,0,6,0,0,0) \\
        (1,10,18,28) & (10,46,253) & 576 & 6 &  (6,0,0,0,12,0,0) \\
                       & (23,59,343) & 1044 & 1 &  (1,0,0,0,2,0,0) \\
        (1,10,22,22) & (7,127,467) & 1088 & 20 &  (20,0,1,0,4,0,0) \\
                       & (17,89,387) & 1504 & 22 &  (22,0,139,153,130,0,0) \\
                       & (7,55,263) & 1148 & 19 &  (19,0,1227,0,886,0,0) \\
                       & (19,67,323) & 1804 & 34 &  (34,0,0,3,0,0,0) \\
                       & (19,67,343) & 1876 & 35 &  (35,0,0,39,0,0,0) \\
                       & (22,58,321) & 2256 & 30 &  (30,0,0,2,0,0,0) \\
                       & (13,37,243) & 1632 & 29 &  (29,0,8,0,14,0,0) \\
                       & (25,49,303) & 2256 & 44 &  (44,0,0,22,0,0,0) \\
                       & (27,51,319) & 2256 & 31 &  (31,0,31,20,3,0,0) \\
                       & (41,41,323) & 2948 & 36 &  (36,0,0,6,2,0,0) \\
        (1,12,12,40) & (25,85,423) & 948 & 7 &  (7,0,1,0,1,0,0) \\
        (1,14,14,22) & (23,23,225) & 952 & 9 &  (9,0,13,0,8,0,0) \\
                       & (31,31,273) & 1352 & 13 &  (13,0,0,2,0,0,0) \\
        (1,16,16,16) & (11,101,401) & 1230 & 12 &  (12,0,219,0,35,0,15) \\
                       & (16,64,325) & 2196 & 13 &  (13,0,2,2,0,0,0) \\
                       & (8,44,227) & 1512 & 14 &  (14,0,88,0,88,0,0) \\
                       & (21,57,301) & 2196 & 16 &  (16,0,0,40,0,0,0) \\
                       & (20,32,243) & 1512 & 10 &  (10,0,5,0,0,0,0) \\
        (2,3,22,118) & (41,77,463) & 2160 & 4 &  (4,0,0,0,4,0,0) \\
         (2,3,28,58) & (17,101,463) & 1408 & 6 &  (6,0,0,0,80,0,0) \\
                       & (17,101,463) & 1472 & 8 &  (8,0,0,0,86,0,0) \\
                       & (39,87,475) & 2552 & 1 &  (1,0,0,0,1,7,0) \\
                       & (39,87,475) & 2668 & 2 &  (2,0,0,0,2,0,0) \\
         (2,3,38,38) & (11,131,533) & 1200 & 7 &  (7,0,0,0,215,0,0) \\
                       & (23,71,389) & 2616 & 8 &  (8,0,0,1,0,0,0) \\
         (2,4,12,82) & (58,34,371) & 2480 & 12 &  (12,0,0,0,4,0,0) \\
                       & (33,39,325) & 1438 & 28 &  (28,0,9,0,71,0,0) \\
                       & (33,39,325) & 1460 & 11 &  (11,0,8,0,6,0,0) \\
                       & (39,33,325) & 1678 & 27 &  (27,0,16,0,32,0,0) \\
                       & (39,33,325) & 1724 & 15 &  (15,0,2,0,4,0,0) \\
                       & (42,48,357) & 1798 & 29 &  (29,0,0,0,19,0,0) \\
                       & (42,48,357) & 1856 & 9 &  (9,0,0,0,12,0,0) \\
                       & (48,42,357) & 2048 & 19 &  (19,0,5,0,1,0,1) \\
                       & (48,42,357) & 2110 & 4 &  (4,0,17,0,53,0,1) \\
                       & (51,57,447) & 2180 & 13 &  (13,0,0,0,12,0,17) \\
                       & (51,57,447) & 2230 & 6 &  (6,0,0,0,2,0,0) \\
                       & (45,45,350) & 2048 & 8 &  (8,0,0,0,2,0,0) \\
         (2,4,13,58) & (37,61,413) & 1540 & 1 &  (1,0,0,0,5,64,0) \\
         (2,4,14,46) & (19,67,339) & 864 & 24 &  (24,0,0,0,13,0,0) \\
                       & (20,56,308) & 1152 & 25 &  (25,0,173,0,359,0,0) \\
                       & (21,57,321) & 1152 & 23 &  (23,0,0,0,841,0,0) \\
                       & (26,62,343) & 1440 & 18 &  (18,0,21,0,165,4,0) \\
                       & (29,65,407) & 1440 & 1 &  (1,0,24,0,26,24,0) \\
                       & (56,20,308) & 2640 & 21 &  (21,0,0,0,9,0,0) \\
                       & (62,26,343) & 2796 & 7 &  (7,0,1,0,0,0,0) \\
                       & (20,44,299) & 864 & 26 &  (26,0,793,0,1151,0,0) \\
                       & (34,58,367) & 1728 & 16 &  (16,0,6,0,116,0,8) \\
                       & (25,37,287) & 1152 & 8 &  (8,0,318,0,887,0,0) \\
                       & (28,40,309) & 1440 & 10 &  (10,0,6,0,266,0,3) \\
                       & (35,47,329) & 1440 & 15 &  (15,0,3,5,11,0,0) \\
                       & (37,25,287) & 1896 & 19 &  (19,0,3,0,1,0,0) \\
         (2,4,16,34) & (16,100,440) & 1000 & 12 &  (12,0,0,0,817,0,0) \\
                       & (13,73,339) & 800 & 38 &  (38,0,0,0,19,0,0) \\
                       & (13,73,339) & 820 & 44 &  (44,0,1,0,182,0,0) \\
                       & (15,69,339) & 962 & 43 &  (43,0,28,0,1999,0,0) \\
                       & (15,69,339) & 1036 & 11 &  (11,0,15,0,2086,0,0) \\
                       & (21,75,421) & 1180 & 47 &  (47,0,84,0,861,76,0) \\
                       & (21,75,421) & 1250 & 19 &  (19,0,0,0,184,50,0) \\
                       & (12,54,297) & 770 & 41 &  (41,0,2964,0,4549,0,0) \\
                       & (12,54,297) & 796 & 45 &  (45,0,1248,0,1808,0,0) \\
                       & (24,66,353) & 1360 & 15 &  (15,0,26,0,209,352,24) \\
                       & (24,66,353) & 1394 & 20 &  (20,0,17,0,70,6,0) \\
                       & (20,56,306) & 1040 & 36 &  (36,0,1768,0,2253,0,0) \\
                       & (20,56,306) & 1120 & 10 &  (10,0,504,0,876,0,0) \\
                       & (26,62,339) & 1144 & 48 &  (48,0,80,18,412,68,0) \\
                       & (26,62,339) & 1232 & 17 &  (17,0,4,0,261,14,0) \\
                       & (15,45,285) & 914 & 42 &  (42,0,5345,0,7519,0,0) \\
                       & (15,45,285) & 976 & 13 &  (13,0,1281,0,1855,0,0) \\
                       & (23,53,307) & 1120 & 49 &  (49,0,115,10,266,4,2) \\
                       & (23,53,307) & 1202 & 18 &  (18,0,274,2,803,0,1) \\
                       & (22,46,309) & 1300 & 46 &  (46,0,112,10,511,24,0) \\
                       & (22,46,309) & 1400 & 16 &  (16,0,455,0,931,0,39) \\
                       & (31,55,330) & 1768 & 14 &  (14,0,0,14,0,0,0) \\
                       & (31,55,330) & 1904 & 1 &  (1,0,12,2,23,0,0) \\
         (2,4,18,28) & (26,44,293) & 854 & 27 &  (27,0,0,0,7,0,0) \\
                       & (26,44,293) & 976 & 9 &  (9,0,0,0,15,0,0) \\
                       & (44,26,293) & 1024 & 11 &  (11,0,8,0,16,0,0) \\
                       & (44,26,293) & 1166 & 4 &  (4,0,3,0,8,0,0) \\
                       & (25,37,279) & 736 & 18 &  (18,0,0,0,12,0,0) \\
                       & (24,30,261) & 686 & 28 &  (28,0,34,0,72,0,0) \\
                       & (24,30,261) & 724 & 19 &  (19,0,31,0,27,0,0) \\
                       & (30,24,261) & 830 & 29 &  (29,0,33,0,44,0,0) \\
                       & (30,24,261) & 940 & 15 &  (15,0,1,0,4,0,0) \\
                       & (36,42,327) & 1060 & 21 &  (21,0,0,0,1,0,0) \\
                       & (42,36,327) & 1276 & 17 &  (17,0,0,0,0,0,1) \\
                       & (31,31,270) & 896 & 22 &  (22,0,0,0,0,0,0) \\
                       & (31,31,270) & 1024 & 8 &  (8,0,2,0,8,0,0) \\
                       & (39,39,313) & 1036 & 10 &  (10,0,0,0,6,0,0) \\
                       & (39,39,313) & 1184 & 1 &  (1,0,0,0,22,0,0) \\
         (2,4,22,22) & (11,131,536) & 864 & 12 &  (12,0,0,0,1292,0,0) \\
                       & (7,103,399) & 648 & 43 &  (43,0,0,0,71,0,0) \\
                       & (14,98,501) & 1080 & 1 &  (1,38,255,154,1123,158,0) \\
                       & (9,81,344) & 864 & 44 &  (44,0,1113,0,1977,0,0) \\
                       & (10,82,361) & 864 & 42 &  (42,0,0,0,404,0,0) \\
                       & (13,85,367) & 1080 & 22 &  (22,0,169,0,1011,168,0) \\
                       & (16,88,393) & 1296 & 20 &  (20,0,88,68,582,0,61) \\
                       & (8,68,329) & 648 & 45 &  (45,0,5089,0,7025,0,0) \\
                       & (14,74,341) & 1080 & 18 &  (18,0,497,154,1704,0,0) \\
                       & (10,58,309) & 864 & 11 &  (11,0,3162,0,7156,0,0) \\
                       & (13,61,335) & 1080 & 13 &  (13,16,2235,231,2695,81,152) \\
                       & (21,69,344) & 1728 & 16 &  (16,0,147,150,147,0,0) \\
                       & (20,56,313) & 1788 & 48 &  (48,0,138,0,85,0,0) \\
                       & (16,40,261) & 1416 & 46 &  (46,0,120,0,251,0,3) \\
                       & (19,43,279) & 1476 & 49 &  (49,0,274,0,173,0,1) \\
                       & (28,52,341) & 2160 & 2 &  (2,0,0,5,22,56,0) \\
                       & (52,28,341) & 3888 & 35 &  (35,0,0,0,2,0,0) \\
                       & (20,32,261) & 1668 & 17 &  (17,0,383,1,319,1,0) \\
                       & (32,44,313) & 2412 & 30 &  (30,0,0,1,8,0,0) \\
                       & (23,23,247) & 1356 & 21 &  (21,0,0,0,4,0,0) \\
                       & (31,31,268) & 1968 & 47 &  (47,0,43,0,73,0,0) \\
                       & (31,31,279) & 2460 & 19 &  (19,0,0,0,1,4,0) \\
                       & (32,32,273) & 2040 & 28 &  (28,0,27,0,65,0,0) \\
                       & (32,32,273) & 2592 & 50 &  (50,0,1,0,0,0,0) \\
                       & (39,39,311) & 2100 & 31 &  (31,0,0,0,8,0,3) \\
         (2,5,12,26) & (8,80,341) & 528 & 8 &  (8,0,0,0,32,0,0) \\
                       & (23,59,327) & 780 & 1 &  (1,0,60,0,252,82,0) \\
                       & (23,59,327) & 858 & 2 &  (2,0,0,0,5,0,0) \\
          (2,6,8,38) & (14,50,289) & 720 & 9 &  (9,0,0,0,530,0,0) \\
                       & (21,57,333) & 1080 & 15 &  (15,0,26,0,14,0,0) \\
                       & (26,50,313) & 1080 & 12 &  (12,0,35,0,33,0,0) \\
                       & (28,52,331) & 1200 & 16 &  (16,0,1,0,74,0,18) \\
                       & (22,34,265) & 720 & 25 &  (25,0,0,0,447,0,0) \\
                       & (25,37,271) & 1080 & 1 &  (1,0,47,0,42,0,0) \\
                       & (33,33,275) & 1440 & 13 &  (13,0,0,0,7,0,0) \\
         (2,6,10,22) & (14,62,299) & 864 & 56 &  (56,0,150,0,120,0,0) \\
                       & (15,63,307) & 864 & 15 &  (15,0,26,0,34,0,0) \\
                       & (16,64,331) & 864 & 58 &  (58,0,3,0,325,0,0) \\
                       & (10,46,253) & 432 & 13 &  (13,0,0,0,1774,0,0) \\
                       & (16,40,249) & 864 & 22 &  (22,0,76,0,297,0,6) \\
                       & (19,43,255) & 864 & 53 &  (53,0,26,0,36,0,0) \\
                       & (22,46,275) & 864 & 20 &  (20,0,0,0,61,0,0) \\
                       & (22,46,319) & 1152 & 38 &  (38,0,0,0,8,0,0) \\
                       & (13,25,215) & 864 & 55 &  (55,0,0,0,2,0,0) \\
                       & (15,27,239) & 864 & 21 &  (21,0,20,0,18,0,0) \\
                       & (17,29,223) & 1008 & 30 &  (30,0,5,0,0,0,0) \\
                       & (18,30,223) & 1080 & 52 &  (52,0,1,0,7,0,0) \\
                       & (20,32,241) & 1080 & 54 &  (54,0,5,0,59,0,0) \\
                       & (20,32,261) & 1008 & 29 &  (29,0,30,0,60,0,0) \\
                       & (22,34,253) & 1152 & 32 &  (32,0,10,0,55,0,0) \\
                       & (26,38,283) & 1080 & 18 &  (18,0,1,0,2,0,0) \\
                       & (29,17,223) & 1152 & 39 &  (39,0,0,0,11,0,0) \\
                       & (32,20,241) & 1512 & 16 &  (16,0,1,0,5,0,0) \\
                       & (32,20,261) & 1440 & 4 &  (4,0,1,0,1,0,0) \\
                       & (38,26,275) & 1512 & 1 &  (1,0,0,0,3,0,0) \\
                       & (21,21,235) & 864 & 51 &  (51,0,25,0,10,0,0) \\
                       & (25,25,243) & 1008 & 36 &  (36,0,0,0,1,0,0) \\
                       & (25,25,243) & 1152 & 33 &  (33,0,0,0,21,0,0) \\
                       & (28,28,253) & 1008 & 35 &  (35,0,0,0,16,0,0) \\
                       & (28,28,253) & 1440 & 6 &  (6,0,0,0,17,0,0) \\
                       & (31,31,259) & 1440 & 40 &  (40,0,0,0,7,0,0) \\
                       & (33,33,273) & 1728 & 23 &  (23,0,0,0,1,0,0) \\
         (2,6,14,14) & (8,80,319) & 768 & 64 &  (64,0,44,0,73,0,1) \\
                       & (10,82,355) & 768 & 73 &  (73,0,109,0,385,0,0) \\
                       & (11,83,389) & 768 & 14 &  (14,0,7,0,34,0,0) \\
                       & (9,57,273) & 768 & 60 &  (60,0,204,0,806,0,18) \\
                       & (10,58,271) & 768 & 22 &  (22,0,123,0,422,0,4) \\
                       & (14,62,283) & 768 & 19 &  (19,0,4,0,99,0,0) \\
                       & (14,62,371) & 1032 & 17 &  (17,0,77,0,21,0,0) \\
                       & (13,49,277) & 1032 & 71 &  (71,0,56,0,68,0,0) \\
                       & (13,49,277) & 1032 & 80 &  (80,0,56,0,68,0,0) \\
                       & (8,32,215) & 768 & 67 &  (67,0,12,0,52,0,0) \\
                       & (9,33,233) & 768 & 21 &  (21,0,154,0,197,0,0) \\
                       & (10,34,251) & 768 & 62 &  (62,0,105,0,0,0,0) \\
                       & (11,35,217) & 768 & 65 &  (65,0,24,0,23,0,0) \\
                       & (13,37,233) & 768 & 66 &  (66,0,7,0,37,0,0) \\
                       & (13,37,253) & 1032 & 63 &  (63,0,49,0,75,0,0) \\
                       & (15,39,241) & 768 & 69 &  (69,0,39,0,33,0,0) \\
                       & (15,39,253) & 768 & 68 &  (68,0,36,0,8,0,0) \\
                       & (16,40,263) & 1032 & 61 &  (61,0,40,0,77,0,0) \\
                       & (17,41,265) & 1272 & 28 &  (28,0,19,0,13,0,0) \\
                       & (18,42,263) & 1272 & 25 &  (25,0,0,0,7,0,0) \\
                       & (20,44,255) & 1536 & 82 &  (82,0,1,0,2,0,0) \\
                       & (21,45,273) & 1536 & 23 &  (23,0,8,0,14,0,0) \\
                       & (18,30,221) & 1032 & 72 &  (72,0,18,0,389,0,0) \\
                       & (21,33,243) & 1272 & 32 &  (32,0,18,0,9,0,0) \\
                       & (26,38,271) & 1272 & 81 &  (81,0,0,0,49,0,0) \\
                       & (26,38,271) & 1272 & 86 &  (86,0,0,0,49,0,0) \\
                       & (23,23,225) & 1536 & 74 &  (74,0,8,0,8,0,0) \\
                       & (23,23,229) & 1032 & 16 &  (16,0,16,0,20,0,0) \\
                       & (25,25,265) & 1272 & 18 &  (18,0,0,0,34,0,0) \\
                       & (34,34,347) & 1272 & 1 &  (1,0,0,0,12,0,8) \\
         (2,7,10,16) & (31,31,271) & 544 & 9 &  (9,0,0,0,1,0,0) \\
          (2,8,8,18) & (7,79,325) & 468 & 34 &  (34,0,0,0,86,0,0) \\
                       & (6,72,325) & 528 & 9 &  (9,0,0,0,140,0,0) \\
                       & (12,66,315) & 648 & 37 &  (37,0,53,0,379,52,0) \\
                       & (12,66,315) & 702 & 17 &  (17,0,10,0,84,4,0) \\
                       & (6,48,249) & 414 & 33 &  (33,0,2103,0,2548,0,0) \\
                       & (6,48,249) & 456 & 35 &  (35,0,345,0,344,0,0) \\
                       & (13,49,249) & 504 & 28 &  (28,0,1194,0,3295,0,0) \\
                       & (13,49,249) & 576 & 8 &  (8,0,649,0,1683,0,0) \\
                       & (13,49,259) & 756 & 36 &  (36,0,394,8,781,114,0) \\
                       & (13,49,259) & 864 & 14 &  (14,0,562,1,739,0,6) \\
                       & (21,57,313) & 588 & 10 &  (10,0,55,0,126,0,0) \\
                       & (21,57,313) & 672 & 1 &  (1,0,0,0,187,8,0) \\
                       & (16,46,257) & 576 & 11 &  (11,0,101,0,113,0,0) \\
                       & (16,46,257) & 654 & 15 &  (15,0,127,0,203,0,0) \\
                       & (28,34,257) & 1134 & 16 &  (16,0,0,0,11,4,0) \\
                       & (28,34,257) & 1296 & 2 &  (2,0,22,0,10,0,0) \\
        (2,10,10,10) & (3,105,381) & 320 & 44 &  (44,0,0,0,67,0,0) \\
                       & (3,105,381) & 328 & 57 &  (57,0,0,0,377,0,0) \\
                       & (7,91,371) & 640 & 54 &  (54,0,58,0,62,0,0) \\
                       & (7,91,371) & 656 & 11 &  (11,0,37,0,169,0,0) \\
                       & (3,69,277) & 320 & 46 &  (46,0,225,0,764,0,0) \\
                       & (3,69,277) & 328 & 56 &  (56,0,771,0,1060,0,0) \\
                       & (7,67,279) & 640 & 52 &  (52,0,100,0,148,0,3) \\
                       & (7,67,279) & 656 & 63 &  (63,0,85,11,475,0,0) \\
                       & (9,63,265) & 984 & 26 &  (26,0,1,0,1,0,0) \\
                       & (11,59,319) & 824 & 13 &  (13,0,12,0,0,0,0) \\
                       & (11,59,319) & 832 & 55 &  (55,0,18,0,0,0,0) \\
                       & (7,43,231) & 824 & 64 &  (64,0,21,0,49,0,0) \\
                       & (7,43,231) & 832 & 51 &  (51,0,94,0,23,0,0) \\
                       & (9,45,243) & 824 & 65 &  (65,0,265,0,651,0,1) \\
                       & (9,45,243) & 832 & 53 &  (53,0,98,14,85,0,0) \\
                       & (13,49,251) & 1112 & 66 &  (66,0,22,1,76,0,0) \\
                       & (13,49,251) & 1120 & 18 &  (18,0,65,4,72,2,0) \\
                       & (15,51,271) & 1312 & 16 &  (16,0,1,0,3,0,0) \\
                       & (9,33,231) & 640 & 50 &  (50,0,438,0,421,0,0) \\
                       & (9,33,231) & 656 & 62 &  (62,0,160,0,188,0,0) \\
                       & (17,41,247) & 1112 & 67 &  (67,0,0,0,10,0,0) \\
                       & (17,41,247) & 1120 & 21 &  (21,0,9,0,11,0,0) \\
                       & (9,27,193) & 984 & 25 &  (25,0,1,0,1,0,0) \\
                       & (13,25,207) & 640 & 48 &  (48,0,1,0,0,0,0) \\
                       & (17,29,215) & 824 & 12 &  (12,0,138,0,172,0,0) \\
                       & (17,29,215) & 832 & 49 &  (49,0,138,0,516,0,0) \\
                       & (19,31,231) & 1648 & 19 &  (19,0,0,0,2,0,0) \\
                       & (19,31,231) & 1664 & 59 &  (59,0,0,0,15,0,0) \\
                       & (19,31,235) & 1112 & 15 &  (15,0,72,0,56,0,0) \\
                       & (19,31,235) & 1120 & 24 &  (24,0,10,0,19,0,0) \\
                       & (27,39,299) & 1112 & 14 &  (14,0,0,0,32,0,0) \\
                       & (27,39,299) & 1120 & 1 &  (1,0,6,0,62,0,0) \\
                       & (35,23,243) & 2224 & 22 &  (22,0,0,0,5,0,0) \\
                       & (35,23,243) & 2240 & 2 &  (2,0,0,0,1,0,0) \\
           (3,8,8,8) & (11,47,283) & 880 & 9 &  (9,0,0,0,0,0,0) \\
                       & (13,25,213) & 1120 & 12 &  (12,0,0,0,3,0,0) \\
          (4,4,6,22) & (9,57,289) & 546 & 37 &  (37,0,0,0,7,0,0) \\
                       & (13,61,289) & 330 & 12 &  (12,0,0,0,544,0,0) \\
                       & (6,42,223) & 354 & 30 &  (30,0,1147,0,2200,0,0) \\
                       & (12,48,255) & 510 & 11 &  (11,0,0,0,497,110,0) \\
                       & (12,48,256) & 408 & 28 &  (28,0,0,0,78,0,0) \\
                       & (9,33,211) & 438 & 8 &  (8,0,612,0,621,0,0) \\
                       & (13,37,224) & 600 & 10 &  (10,0,312,2,366,0,0) \\
                       & (17,41,243) & 570 & 1 &  (1,0,30,21,105,4,0) \\
                       & (19,43,261) & 534 & 14 &  (14,0,54,0,160,0,0) \\
                       & (19,43,276) & 840 & 15 &  (15,0,0,2,0,0,0) \\
                       & (21,45,283) & 420 & 35 &  (35,0,2,0,3,0,0) \\
                       & (16,28,219) & 348 & 13 &  (13,0,289,0,212,0,0) \\
                       & (18,30,221) & 402 & 34 &  (34,0,182,0,485,0,0) \\
                       & (24,36,261) & 954 & 43 &  (43,0,0,0,0,0,3) \\
                       & (31,31,283) & 588 & 17 &  (17,0,10,0,8,0,0) \\
          (4,4,7,16) & (11,59,283) & 288 & 14 &  (14,0,0,0,4,0,0) \\
                       & (14,50,261) & 432 & 15 &  (15,0,5,0,1,0,0) \\
                       & (15,39,233) & 720 & 13 &  (13,0,0,3,0,0,0) \\
                       & (17,41,283) & 864 & 5 &  (5,0,0,0,22,0,0) \\
                       & (20,32,285) & 1008 & 18 &  (18,0,0,0,21,0,0) \\
                       & (23,23,217) & 1008 & 12 &  (12,0,0,3,0,0,0) \\
         (4,4,10,10) & (4,70,279) & 380 & 76 &  (76,0,0,0,16,0,0) \\
                       & (9,69,319) & 308 & 67 &  (67,0,0,0,1,0,0) \\
                       & (8,62,263) & 430 & 22 &  (22,0,0,14,67,0,0) \\
                       & (10,64,299) & 406 & 16 &  (16,0,56,0,34,0,0) \\
                       & (5,53,232) & 440 & 77 &  (77,0,1070,0,1264,0,0) \\
                       & (9,57,249) & 512 & 74 &  (74,0,218,0,273,0,0) \\
                       & (9,57,252) & 640 & 21 &  (21,0,94,32,73,4,0) \\
                       & (4,46,219) & 332 & 14 &  (14,0,1324,0,1409,0,0) \\
                       & (6,48,223) & 296 & 66 &  (66,0,712,0,285,0,0) \\
                       & (7,43,215) & 260 & 79 &  (79,0,1237,0,587,0,0) \\
                       & (13,49,247) & 448 & 80 &  (80,0,15,0,5,0,0) \\
                       & (10,40,219) & 550 & 81 &  (81,0,0,0,59,0,0) \\
                       & (10,40,223) & 744 & 86 &  (86,0,0,0,3,0,0) \\
                       & (15,39,321) & 1056 & 68 &  (68,0,0,0,35,0,0) \\
                       & (14,20,195) & 888 & 101 &  (101,0,0,0,1,0,0) \\
          (4,6,6,10) & (14,38,229) & 416 & 41 &  (41,0,0,0,1,0,1) \\
                       & (12,24,197) & 320 & 50 &  (50,0,0,0,1,0,0) \\
                       & (24,24,225) & 448 & 21 &  (21,0,0,0,3,0,0) \\
           (6,6,6,6) & (5,69,267) & 400 & 7 &  (7,0,4,0,14,0,0) \\
                       & (3,59,223) & 368 & 37 &  (37,0,142,0,226,0,0) \\
                       & (3,43,207) & 400 & 58 &  (58,0,30,0,30,0,0) \\
                       & (7,47,215) & 736 & 55 &  (55,0,0,0,6,0,0) \\
                       & (5,37,203) & 368 & 38 &  (38,0,63,0,100,0,0) \\
                       & (9,41,211) & 800 & 18 &  (18,0,2,0,18,0,2) \\
                       & (11,27,199) & 736 & 56 &  (56,0,0,0,6,0,0) \\
                       & (13,29,203) & 736 & 47 &  (47,0,0,0,4,0,0) \\
                       & (15,23,199) & 1136 & 68 &  (68,0,2,0,2,0,0) \\
                       & (17,25,203) & 1136 & 69 &  (69,0,0,0,1,0,0) \\
       (1,2,2,10,10) & (5,29,235) & 224 & 19 &  (19,0,2,0,0,0,0) \\
        (1,2,4,4,10) & (5,53,248) & 112 & 76 &  (76,0,1,0,1,0,0) \\
                       & (5,41,212) & 106 & 75 &  (75,0,1,0,1,0,0) \\
                       & (6,30,196) & 106 & 65 &  (65,0,25,0,40,0,0) \\
                       & (11,35,213) & 160 & 67 &  (67,0,0,30,0,0,0) \\
         (1,4,4,4,4) & (4,40,213) & 246 & 58 &  (58,0,0,5,0,0,0) \\
                       & (13,13,200) & 240 & 89 &  (89,0,0,0,1,0,0) \\
        (2,2,2,4,10) & (9,45,251) & 208 & 202 &  (202,0,0,0,2,0,0) \\
                       & (9,45,251) & 320 & 87 &  (87,0,19,0,37,0,0) \\
                       & (13,25,219) & 320 & 101 &  (101,0,14,0,14,0,0) \\
                       & (15,27,245) & 416 & 213 &  (213,1,0,0,0,0,0) \\
                       & (21,9,211) & 272 & 79 &  (79,0,0,0,1,0,0) \\
                       & (21,9,211) & 272 & 109 &  (109,0,0,0,1,0,0) \\
                       & (21,9,211) & 416 & 224 &  (224,0,0,0,1,0,0) \\
         (2,2,2,6,6) & (3,51,235) & 288 & 106 &  (106,0,31,0,24,0,0) \\
                       & (9,33,223) & 384 & 149 &  (149,0,0,0,1,0,0) \\
                       & (9,33,223) & 384 & 189 &  (189,0,0,0,1,0,0) \\
                       & (17,17,215) & 576 & 135 &  (135,0,0,0,2,0,0) \\
                       & (17,17,215) & 576 & 157 &  (157,0,0,0,2,0,0) \\
                       & (17,17,215) & 576 & 285 &  (285,0,0,0,2,0,0) \\
      (1,1,1,1,7,16) & (23,23,217) & 72 & 32 &  (32,0,0,2,0,0,0) 

%% file: chiral_types_table.inc
0&0&0&0&10564\\
1&-3&3&0&32\\
1&-9&3&6&1\\
1&-9&9&0&22\\
2&0&0&0&49661\\
3&-3&-1&4&141\\
3&-3&-3&6&24\\
3&-3&1&2&240\\
3&-3&3&0&740\\
3&-9&-3&12&24\\
3&-9&3&6&95\\
3&-9&5&4&1\\
3&-9&9&0&116\\
4&0&0&0&116304\\
5&-3&1&2&2\\
5&-3&3&0&1507\\
5&-9&9&0&46\\